\documentclass[nonblindrev]{informs3} 

\usepackage{balance}
\usepackage{amsmath}
\usepackage[tight,footnotesize]{subfigure}
\usepackage{graphicx}
\usepackage{bm}
\usepackage{algorithm}
\usepackage[noend]{algorithmic}
\usepackage{multirow}
\usepackage{caption}
\usepackage{amsfonts}

\OneAndAHalfSpacedXII 

\usepackage{natbib}
 \bibpunct[, ]{(}{)}{,}{a}{}{,}%
 \def\bibfont{\small}%
\TheoremsNumberedThrough     

\EquationsNumberedThrough    

\MANUSCRIPTNO{2015} 

\begin{document}


\RUNAUTHOR{Tang et al.}

\RUNTITLE{Robust Adaptive Submodular Maximization}

\TITLE{Robust Adaptive Submodular Maximization}

\ARTICLEAUTHORS{%
\AUTHOR{Shaojie Tang}
\AFF{Naveen Jindal School of Management, The University of Texas at Dallas}
} 

\ABSTRACT{The goal of a sequential decision making problem is to design an interactive policy that adaptively selects a group of items, each selection is based on the feedback from the past, in order to maximize the expected utility of selected items. It has been shown that the utility functions of many real-world applications are adaptive submodular. However, most of existing studies on adaptive submodular optimization focus on the average-case, i.e.,  their objective is to find a policy that maximizes the expected utility over a known distribution of realizations. Unfortunately, a policy that has a good average-case performance may have very poor performance under the worst-case realization. In this study, we propose to study two variants of adaptive submodular optimization problems, namely, worst-case adaptive submodular maximization and robust submodular maximization. The first problem aims to find a policy that maximizes the worst-case utility and the latter one aims to find a policy, if any, that achieves both near optimal average-case utility and worst-case utility simultaneously. We introduce a new class of stochastic functions, called \emph{worst-case submodular function}. For the worst-case adaptive submodular maximization problem subject to a $p$-system constraint, we develop an adaptive worst-case greedy policy that achieves a $\frac{1}{p+1}$ approximation ratio against the optimal worst-case utility if the utility function is worst-case submodular. For the robust adaptive submodular maximization problem subject to cardinality constraints (resp. partition matroid constraints),   if the utility function is both worst-case submodular and adaptive submodular, we develop a hybrid adaptive policy that achieves an approximation close to $1-e^{-\frac{1}{2}}$ (resp. $1/3$) under both worst- and average-case settings simultaneously.  We also describe several applications of our theoretical results, including pool-base active learning, stochastic submodular set cover and adaptive viral marketing. }


\maketitle
\section{Introduction}
Maximizing a submodular function subject to practical constraints has been extensively studied in the literature \citep{tang2020influence,yuan2017no,krause2007near,leskovec2007cost,badanidiyuru2014fast,mirzasoleiman2016fast,ene2018towards,mirzasoleiman2015lazier}. Many machine learning and AI tasks such as viral marketing \citep{golovin2011adaptive}, data summarization \citep{badanidiyuru2014streaming} and active learning \citep{golovin2011adaptive} can be formulated as a submodular maximization problem. While most of existing studies focus on non-adaptive submodular maximization by assuming a deterministic utility function, \cite{golovin2011adaptive} extends this study to the adaptive setting where the utility function is stochastic. Concretely, the input of an adaptive optimization problem is a set of items, and each item is in a particular state drawn from some known prior distribution. There is an  utility function which is defined over items and their states. One must select an item before revealing its realized state.  In the context of experimental design, each test (e.g., blood pressure) is an item and the outcome of a test can be regarded as its state (e.g., possible outcomes of a blood pressure test is low or high). Our objective is to adaptively select a group of items (e.g., a group of tests in experimental design), each selection is based on the feedback from the past, to maximize the average-case utility over the distribution of realizations. Note that a policy that has a good average-case performance may perform poorly under the worst-case realization. This raises our first research question:

\emph{Is it possible to design a policy  that maximizes the worst-case utility?}

Moreover, even if we can find such a policy, this worst-case guarantee often comes at the expense of degraded average-case performance. This raises our second research question:

\emph{Is it possible to design a policy, if any, that achieves both good average-case and worst-case performance simultaneously?}

In this paper, we provide affirmative answers to both questions by studying two variants of adaptive submodular optimization problems: \emph{worst-case adaptive submodular maximization} and \emph{robust submodular maximization}.   In the first problem, our goal is to  find a policy that maximizes the utility under the worst-case realization. The second problem aims to find a policy that achieves both near optimal average-case utility and worst-case utility simultaneously. To tackle these two problems, we introduce a new class of stochastic functions, called \emph{worst-case submodular function}, and we show that this property can be found in a wide range of real-world applications, including pool-based active learning \citep{golovin2011adaptive} and adaptive viral marketing \citep{golovin2011adaptive}.

We first study the worst-case adaptive submodular maximization problem subject to a $p$-system constraint, and develop an adaptive worst-case greedy policy that achieves a $\frac{1}{p+1}$ approximation ratio against the optimal worst-case utility if the utility function is worst-case submodular. Note that the $p$-system  constraint is general enough to subsume many practical constraints, including cardinality, matroid, intersection of $p$ matroids, $p$-matchoid and $p$-extendible constraints, as special cases. We also show that both the approximation ratio and the running time can be improved  for the case of a single cardinality constraint. Then we initiate the study of robust adaptive submodular maximization problem. If the utility function is both worst-case submodular and adaptive submodular, we develop hybrid adaptive policies that achieve nearly $1-e^{-\frac{1}{2}}$ and  $1/3$ approximation subject to cardinality constraints and partition matroid constraints, respectively, under both worst and average cases simultaneously.

\section{Related Works}
\citet{golovin2011adaptive}  introduce the problem of adaptive submodular maximization, where they extend the notion of submodularity and monotonicity to the adaptive setting by introducing \emph{adaptive submodularity} and \emph{adaptive monotonicity}. They develop  a simple adaptive greedy algorithm that achieves a $1-1/e$ approximation for maximizing a monotone adaptive submodular function subject to a cardinality constraint in average case. In \citep{golovin2011adaptive1}, they further extend their results to a $p$-system constraint. When the utility function is not adaptive monotone, \cite{tang2021beyond,tang2021beyond2} develops the first constant factor approximation algorithm in average case. Other variants of this problem have been studied in \citep{tang2021non,tang2022group,tang2022optimal,tang2021adaptive,tang2022partial}.
While most of existing studies on adaptive submodular maximization focus on the average case setting, \cite{guillory2010interactive,golovin2011adaptive} studied the worst-case min-cost submodular set cover problem, their objective, which is different from ours, is to  adaptively select a group of cheapest items  until the resulting utility function achieves some threshold. \cite{cuong2014near} studied the pointwise submodular maximization problem subject to a cardinality constraint in worst case setting. They claimed that if the utility function is pointwise submodular, then a greedy policy achieves a constant approximation ratio in worst case setting. Unfortunately, we show in Section \ref{sec:doesnot} that this result does not hold in general. In particular, we construct a counter example to show that the performance of the aforementioned greedy policy is arbitrarily bad even if the utility function is pointwise submodular. To tackle this problem, we introduce the notation of worst-case submodularity and develop a series of effective solutions for maximizing a worst-case submodular function subject to a $p$-system constraint. Our results are not restricted to any particular applications. Moreover, perhaps surprisingly, we propose the first algorithm that achieves good approximation ratios in both average case and worst case settings simultaneously.

\section{Preliminaries}
In the rest of this paper, we use  $[m]$ to denote the set $\{1, 2, \cdots, m\}$.

\subsection{Items and States.} We consider a set  $E$ of $n$ items (e.g., tests in experimental design), and each item $e \in E$ has a random state $\Phi(e)\in O$ (e.g., possible outcomes of a test) where $O$ represents a set of possible states.  Let a function $\phi: E\rightarrow O$ denote a \emph{realization}, where for each $e\in E$, $\phi(e)$ represents the realization of $\Phi(e)$. In the example of experimental design,  an item $e$ may represent a test, such as the temperature, and
$\Phi(e)$ is the outcome of the test, such as, \emph{high}. The realization of $\Phi$ is unknown initially and one must pick an item before revealing its realized state. There is a known prior probability distribution $p(\phi)=\Pr[\Phi=\phi]$ over $\phi\in U$ where $U$ denotes the set of all realizations. Given any $S\subseteq E$, let $\psi: S\rightarrow O$ denote a \emph{partial realization} and $\mathrm{dom}(\psi)=S$ is called the \emph{domain} of $\psi$.
Given a partial  realization $\psi$ and a realization $\phi$, we say $\phi$ is consistent with $\psi$, denoted $\phi \sim \psi$, if they are equal everywhere in $\mathrm{dom}(\psi)$.  A partial realization $\psi$  is said to be a \emph{subrealization} of  $\psi'$, denoted  $\psi \subseteq \psi'$, if $\mathrm{dom}(\psi) \subseteq \mathrm{dom}(\psi')$ and they are equal everywhere in $\mathrm{dom}(\psi)$.
We use $p(\phi\mid \psi)$ to denote the conditional distribution over realizations conditioned on  a partial realization $\psi$: $p(\phi\mid \psi) =\Pr[\Phi=\phi\mid \Phi\sim \psi ]$. 


\subsection{Policy and Adaptive/Worst-case Submodularity.}
\label{sec:3.2}

We represent a policy using a function $\pi$ that maps a set of partial realizations  to $E$: $\pi: 2^{E}\times O^E \rightarrow E$.  Intuitively, $\pi$ is a mapping from the observations, which are represented as a set of selected items and their realizations, collected so far to the next item to select. 
For example, consider a policy $\pi$, suppose the current observation is $\cup_{e\in S}\{ (e, \Phi(e))\}$ after selecting a set of items $S$ and assume  $\pi(\cup_{e\in S} \{(e, \Phi(e))\})= w$, then $\pi$ selects $w$ as the next item. Note that any randomized policy can be represented as a distribution of a group of deterministic policies, thus we will focus on deterministic policies. 

\begin{definition}\citep{golovin2011adaptive}\emph{[Policy  Concatenation]}
Given two policies $\pi$ and $\pi'$,  let $\pi @\pi'$ denote a policy that runs $\pi$ first, and then runs $\pi'$, ignoring the observation obtained from running $\pi$.
\end{definition}

\begin{definition}\citep{golovin2011adaptive}\emph{[Level-$t$-Truncation of a Policy]}
Given a policy $\pi$, we define its  level-$t$-truncation $\pi_t$  as a policy that runs $\pi$ until it selects $t$ items.
\end{definition}

There is a utility function $f : 2^{E}\times 2^{O} \rightarrow \mathbb{R}_{\geq0}$ from a subset of items and their states to a non-negative real number. Let $E(\pi, \phi)$ denote the subset of items selected by $\pi$ under realization $\phi$. The expected  utility $f_{avg}(\pi)$ of a policy $\pi$ can be written as
\begin{eqnarray}
f_{avg}(\pi)=\mathbb{E}_{\Phi}[f(E(\pi, \Phi), \Phi)]~\nonumber
\end{eqnarray}
 where the expectation is taken over $\Phi$ according to $p(\phi)$. Let $U^+=\{\phi\mid p(\phi)>0\}$ denote the set of all realizations with positive probability. The worst-case  utility $f_{wc}(\pi)$ of a policy $\pi$ can be written as
\begin{eqnarray}
f_{wc}(\pi)=\min_{\phi\in U^+}f(E(\pi, \phi), \phi)~\nonumber
\end{eqnarray}

Let $f_{avg}(e \mid \psi)=\mathbb{E}_{\Phi}[f(\mathrm{dom}(\psi)\cup \{e\}, \Phi)-f(\mathrm{dom}(\psi), \Phi)\mid \Phi \sim \psi]$  denote the conditional expected marginal utility of $e$ on top of a partial realization $\psi$, where the expectation is taken over $\Phi$ with respect to $p(\phi\mid \psi)=\Pr(\Phi=\phi \mid \Phi \sim \psi)$. We next introduce the notations of adaptive submodularity and adaptive monotonicity \citep{golovin2011adaptive}. Intuitively, adaptive submodularity extends the classic notation of submodularity from sets to policies.

\begin{definition}\citep{golovin2011adaptive}\emph{[Adaptive Submodularity and Adaptive Monotonicity]} Consider any two partial realizations $\psi$ and $\psi'$ such that $\psi\subseteq \psi'$. A function   $f: 2^{E}\times O^E\rightarrow \mathbb{R}_{\geq0}$  is called adaptive submodular if for each $e\in E\setminus \mathrm{dom}(\psi')$, we have
\begin{eqnarray}\label{def:33}
f_{avg}(e\mid \psi) \geq f_{avg}(e\mid \psi')
\end{eqnarray}
 A function   $f: 2^{E}\times O^E\rightarrow \mathbb{R}_{\geq0}$ is called adaptive monotone if for all $\psi$ and $e\in E\setminus \mathrm{dom}(\psi)$, we have $f_{avg}(e\mid \psi)\geq 0$.
\end{definition}

By extending the definition of $f$, for any $S\subseteq E$ and partial realization $\psi$, let $f(S, \psi) = \mathbb{E}_{\Phi}[f(S, \Phi)\mid \Phi \sim \psi]$.  We next introduce the worst-case marginal utility $f_{wc}(e \mid \psi)$   of $e$ on top of a partial realization $\psi$.
\begin{eqnarray}
\label{def:wc}
f_{wc}(e \mid \psi)=\min_{o \in O(e, \psi)}\{f(\mathrm{dom}(\psi)\cup\{e\}, \psi\cup\{(e, o)\})-f(\mathrm{dom}(\psi), \psi)\}~\nonumber
\end{eqnarray} where $ O(e, \psi)=\{o\in O\mid \exists \phi: p(\phi\mid\psi)>0, \phi(e)=o\}$ denotes the set of possible states of $e$ conditioned on a partial realization $\psi$.

We next introduce a new class of stochastic functions.
\begin{definition}\emph{[Worst-case Submodularity and Worst-case Monotonicity]} Consider any two partial realizations $\psi$ and $\psi'$ such that $\psi\subseteq \psi'$. A function   $f: 2^{E}\times O^E\rightarrow \mathbb{R}_{\geq0}$  is called worst-case submodular if for each $e\in E\setminus \mathrm{dom}(\psi')$, we have
\begin{eqnarray}\label{def:332}
f_{wc}(e\mid \psi) \geq f_{wc}(e\mid \psi')
\end{eqnarray}
A function $f: 2^{E}\times O^E\rightarrow \mathbb{R}_{\geq0}$ is called worst-case monotone if for each partial realization $\psi$ and $e\in E\setminus \mathrm{dom}(\psi)$, $f_{wc}(e\mid \psi) \geq 0$.
\end{definition}

 Note that  if for each partial realization $\psi$ and $e\in E\setminus \mathrm{dom}(\psi)$, $f_{wc}(e\mid \psi) \geq 0$, then because $f_{avg}(e\mid \psi)=\mathbb{E}_{\Phi}[f(\mathrm{dom}(\psi)\cup \{e\}, \Phi)-f(\mathrm{dom}(\psi), \Phi)\mid \Phi \sim \psi]  \geq \min_{o \in O(e, \psi)}f(\mathrm{dom}(\psi)\cup\{e\}, \psi\cup\{(e, o)\})-f(\mathrm{dom}(\psi), \psi) = f_{wc}(e\mid \psi)$, it is also true that for each partial realization $\psi$ and $e\in E\setminus \mathrm{dom}(\psi)$, $f_{avg}(e\mid \psi) \geq 0$. Therefore, worst-case monotonicity implies adaptive monotonicity. We next introduce the property of minimal dependency \citep{cuong2014near} which states that the utility of any group of items does not depend on the states of any items outside that group.
\begin{definition}\emph{[Minimal Dependency]} For any partial realization $\psi$ and any realization $\phi$ such that $\phi\sim \psi$, we have $f(\mathrm{dom}(\psi), \psi) = f(\mathrm{dom}(\psi), \phi)$. 
\end{definition}

\subsection{Problem Formulation}
Now we are ready to introduce the two optimization problems studied in this paper. We first introduce the notation of independence system.
\begin{definition}[Independence System] Consider a ground set $E$ and a collection of sets $\mathcal{I}\subseteq 2^E$, the pair $(E, \mathcal{I})$ is an \emph{independence system} if it satisfies the following two conditions:
1. The empty set is independent, i.e., $\emptyset \in \mathcal{I}$; 2. $\mathcal{I}$ is downward-closed, that is, $A \in \mathcal{I}$ and $B \subseteq A$ implies that $B \in \mathcal{I}$.
\end{definition}

If $A\in \mathcal{I}$ and $B \subseteq A$ imply that $B = A$, then $B \in \mathcal{I}$ is called  a \emph{base}. Moreover, a set $B \in \mathcal{I}$ is called a base of $R$ if  $B\subseteq R$ and  $B$ is a base of the independence system $(R, 2^R \cap \mathcal{I})$. Let $\mathcal{B}(R)$ denote the collection of all bases of $R$.  We next present the definition of $p$-system.

\begin{definition}[$p$-system] A $p$-system  for an integer $p\geq 1$ is an independence system $(E, \mathcal{I})$ such that for every set
$R\subseteq E$, $p\cdot \min_{B\in \mathcal{B}(R)}|B|\geq  \max_{B\in \mathcal{B}(R)}|B|$.
\end{definition}

In this paper, we study the following two problems.
\paragraph{Worst-case adaptive submodular maximization}
 The worst-case adaptive submodular maximization problem subject to a $p$-system constraint $(E, \mathcal{I})$ can be formulated as follows:
\[\max_{\pi}\{f_{wc}(\pi)\mid E(\pi, \phi)\in \mathcal{I} \mbox{ for all } \phi\in U^+\}\]

\paragraph{Robust adaptive submodular maximization}
The aim of this problem is to find a policy that performs well under both worst case setting and average case setting. We first introduce some notations. Given an independence system $(E, \mathcal{I})$, let
\[\pi_{wc}^*\in \argmax_{\pi}\{f_{wc}(\pi)\mid E(\pi, \phi)\in \mathcal{I} \mbox{ for all } \phi\in U^+\}\] denote the optimal worst-case adaptive policy and let \[\pi_{avg}^*\in \argmax_{\pi}\{f_{avg}(\pi)\mid E(\pi, \phi)\in \mathcal{I} \mbox{ for all } \phi\in U^+\}\] denote the optimal average-case adaptive policy. Define $\alpha(\pi)=\min\{\frac{f_{wc}(\pi)}{f_{wc}(\pi^*_{wc})}, \frac{f_{avg}(\pi)}{f_{avg}(\pi^*_{avg})}\}$ as the \emph{robustness ratio} of a policy $\pi$. Intuitively, a larger $\alpha(\pi)$ indicates that $\pi$ achieves better performance under both average- and worst-case settings. The robust adaptive submodular maximization problem is to find a policy that maximizes $\alpha(\pi)$, i.e.,
\[\max_{\pi}\{\alpha(\pi)\mid E(\pi, \phi)\in \mathcal{I} \mbox{ for all } \phi\in U^+\}\]

\section{Worst-case Adaptive Submodular Maximization}
  We first study the problem of worst-case adaptive submodular maximization. We present an \emph{Adaptive Worst-case Greedy Policy} $\pi^w$ for this problem.
 The detailed implementation of $\pi^w$ is listed in Algorithm \ref{alg:LPP2}. It starts with an empty set and  at each round $t$, $\pi^{w}$  selects an item $e_t$ that maximizes the worst-case marginal utility on top of the current observation $\psi_{t-1}$, i.e.,
 \[e_t \in \argmax_{e\in E: \{e\}\cup \mathrm{dom}(\psi_{t-1}) \in \mathcal{I}} f_{wc}(e\mid \psi_{t-1})\]
After observing the state $\Phi(e_t)$ of $e_t$, update  the current partial realization $\psi_{t}$ using  $\psi_{t-1}\cup\{(e_t,\Phi(e_t))\}$.  This process iterates until the current solution can not be further expanded.

Recall that $\pi^*_{wc}$ denotes the optimal policy, we next show that $\pi^w$ achieves a $\frac{1}{p+1}$ approximation ratio, i.e.,  $f_{wc}(\pi^w) \geq \frac{1}{p+1}f_{wc}(\pi^*_{wc})$, if the utility function is worst-case monotone and worst-case submodular, and it satisfies the property of minimal dependency.

\begin{algorithm}[hptb]
\caption{ Adaptive Worst-case Greedy Policy for $p$-System Constraint $\pi^w$}
\label{alg:LPP2}
\begin{algorithmic}[1]
\STATE $t=1; \psi_0=\emptyset; V = E$.
\WHILE {$V\neq \emptyset$}
\STATE select $e_t \in \argmax_{e\in V} f_{wc}(e\mid \psi_{t-1})$;
\STATE $\psi_{t}\leftarrow \psi_{t-1}\cup\{(e_t, \Phi(e_t))\}$;
\STATE  $V=\{e\in E\mid \{e\}\cup \mathrm{dom}(\psi_{t})\in \mathcal{I}, e\notin\mathrm{dom}(\psi_{t})\}$;
\STATE $t\leftarrow t+1$;
\ENDWHILE
\end{algorithmic}
\end{algorithm}

\begin{theorem}
\label{thm:1}
If the utility function $f: 2^E\times O^E\rightarrow \mathbb{R}_{\geq 0}$  is worst-case monotone, worst-case submodular with respect to $p(\phi)$ and it satisfies the property of minimal dependency, then $f_{wc}(\pi^w) \geq \frac{1}{p+1}f_{wc}(\pi^*_{wc})$ subject to $p$-system constraints.
\end{theorem}
\emph{Proof:} Assume $\phi'$ is the worst-case realization of $\pi^w$, i.e., $\phi'=\argmin_{\phi} f(E(\pi^w, \phi), \phi)$, and it selects $b$ items conditioned on $\phi'$, i.e., $|E(\pi^w, \phi')|=b$. For each $t\in[b]$, let $S_t$ denote the first $t$ items selected by $\pi^w$ conditioned on $\phi'$, and  let $\psi'_{t}$ denote the partial realization of  $S_t$ conditioned on $\phi'$, i.e., $\psi'_{t} = \{(e, \phi'(e))\mid e\in S_t\}$. Thus, $S_b$ is a collection of all items selected by $\pi^w$ and $\psi'_b$ is the partial realization of $S_b$ conditioned on $\phi'$. Let $e'_t$ denote the $t$-th item selected by $\pi^w$ conditioned on $\phi'$, i.e., $S_t=\{e'_1, e'_2, \cdots, e'_t\}$. We first show that $f_{wc}(\pi^w)\geq\sum_{t\in [b]} f_{wc}(e'_t\mid \psi'_{t-1})$.
\begin{eqnarray}
f_{wc}(\pi^w) && = f(E(\pi^w, \phi'), \phi') = f(S_b, \psi'_b) = \sum_{t\in [b]}  \{f(S_{t}, \psi'_{t})-f(S_{t-1}, \psi'_{t-1})\}~\nonumber\\
&&=  \sum_{t\in [b]}  \{f(S_{t-1}\cup \{e'_{t}\}, \psi'_{t-1}\cup\{(e'_{t}, \phi'(e'_{t}))\})-f(S_{t-1}, \psi'_{t-1})\}~\nonumber\\
&&\geq \sum_{t\in [b]}\min_{o\in O(e'_t, \psi'_{t-1})}\{f(S_{t-1}\cup \{e'_{t}\}, \psi'_{t-1}\cup\{(e'_{t}, o)\})-f(S_{t-1}, \psi'_{t-1})\}~\nonumber\\
&&= \sum_{t\in [b]} f_{wc}(e'_t\mid \psi'_{t-1})\label{eq:patch}
\end{eqnarray}
The second equality is due to the assumption of minimal dependency. To prove this theorem, it suffices to demonstrate that for any optimal policy $\pi^*_{wc}$, there exists a realization $\phi^*$  such that $f(E(\pi^*_{wc}, \phi^*), \phi^*) \leq (1+p)f(E(\pi^w, \phi'), \phi')$. The rest of the proof is devoted to constructing such a realization $\phi^*$. First, we ensure that $\phi^*$ is consistent with $\psi'_b$ by setting $\phi^*(e)=\phi'(e)$ for each $e\in \mathrm{dom}(\psi'_b)$. Next, we complete the construction of $\phi^*$ by simulating the execution of  $\pi^*_{wc}$ conditioned on $\psi'_b$. Let $S^*_i$ denote the first $i$ items selected by $\pi^*_{wc}$ during the process of construction and let $\psi^*_i$ denote the partial realization of $S^*_i$. Starting with $i=1$ and let $\psi^*_0=\emptyset$, assume the optimal policy $\pi^*_{wc}$ picks $e^*_i$ as the $i$-th item after observing $\psi^*_i$, we set the state $\phi^*(e^*_i)$ of $e^*_i$ to  $\argmin_{o\in O(e^*_i, \psi'_b\cup \psi^*_{i-1})}\{f(S_b\cup S^*_{i-1}\cup\{e^*_i\}, \psi'_b\cup \psi^*_{i-1}\cup\{(e^*_i, o)\}) - f(S_b\cup S^*_{i-1}, \psi'_b\cup \psi^*_{i-1})\}$, then update the observation using $\psi^*_{i}\leftarrow \psi^*_{i-1}\cup\{(e^*_i, \phi^*(e^*_i))\}$. This construction process continues until $\pi^*_{wc}$ does not select new items. Intuitively, in each round $i$ of $\pi^*_{wc}$, we choose a state for $e^*_i$ to minimize the marginal utility of $e^*_i$ on top of the partial realization $\psi'_b\cup \psi^*_{i-1}$. Assume  $|E(\pi^*_{wc}, \phi^*)|=b^*$, i.e., $\pi^*_{wc}$ selects $b^*$ items conditioned on $\phi^*$, thus, $S^*_{b^*}$ is a collection of all items selected by $\pi^*_{wc}$ and $\psi^*_{b^*}$ is the partial realization of $S^*_{b^*}$ conditioned on $\phi^*$. Note that there may exist multiple realizations which meet the above description, we choose an arbitrary one as $\phi^*$. Then we have
\begin{eqnarray}
f_{wc}(\pi^*_{wc}) && \leq f(E(\pi^*_{wc}, \phi^*), \phi^*) \leq f(E(\pi^*_{wc}@\pi^w, \phi^*), \phi^*) = f(S_b\cup S^*_{b^*}, \psi'_b\cup\psi^*_{b^*}) ~\nonumber\\
&&= f(S_b, \psi'_b) + \sum_{i\in [b^*]} \{ f(S_b\cup S^*_{i}, \psi'_b\cup\psi^*_{i})-f(S_b\cup S^*_{i-1}, \psi'_b\cup\psi^*_{i-1})\}~\nonumber\\
&&= f(S_b, \psi'_b) + \sum_{i\in [b^*]} \{ f(S_b\cup S^*_{i-1}\cup \{e^*_{i}\}, \psi'_b\cup\psi^*_{i-1}\cup\{(e^*_{i}, \phi^*(e^*_{i}))\})-f(S_b\cup S^*_{i-1}, \psi'_b\cup\psi^*_{i-1})\}~\nonumber\\
&&= f(S_b, \psi'_b) + \sum_{i\in [b^*]} f_{wc}(e^*_{i}\mid \psi'_b\cup\psi^*_{i-1}) ~\nonumber\\
&&=  f_{wc}(\pi^w) + \sum_{i\in [b^*]} f_{wc}(e^*_{i}\mid \psi'_b\cup\psi^*_{i-1}) \label{eq:e}
\end{eqnarray}

The second inequality is due to $f: 2^E\times O^E\rightarrow \mathbb{R}_{\geq 0}$  is worst-case monotone. The first and the last equalities are due to the assumption of minimal dependency. The fourth equality is due to $\phi^*(e^*_i)$ is a state of $e^*_i$ minimizing the marginal utility $f(S_b\cup S^*_{i-1}\cup\{e^*_i\}, \psi'_b\cup \psi^*_{i-1}\cup\{(e^*_i, o)\}) - f(S_b\cup S^*_{i-1}, \psi'_b\cup \psi^*_{i-1})$, i.e.,   $\phi^*(e^*_i)=\argmin_{o\in O(e^*_i, \psi'_b\cup \psi^*_{i-1})}\{f(S_b\cup S^*_{i-1}\cup\{e^*_i\}, \psi'_b\cup \psi^*_{i-1}\cup\{(e^*_i, o)\}) - f(S_b\cup S^*_{i-1}, \psi'_b\cup \psi^*_{i-1})\}$.

We next show that for all $t\in[b]$ and $i\in [b^*]$,
\begin{eqnarray}
\label{eq:1}f_{wc}(e^*_{i}\mid \psi'_b\cup\psi^*_{i-1}) \leq  f_{wc}(e^*_{i}\mid \psi'_{t-1})
 \end{eqnarray}
 If $e^*_{i}\notin \mathrm{dom}(\psi'_b\cup\psi^*_{i-1})$, the above inequality holds due to $f$ is worst-case submodular and $\psi'_{t-1}\subseteq \psi'_b\cup\psi^*_{i-1}$ for all $t\in[b]$. If $e^*_{i}\in \mathrm{dom}(\psi'_b\cup\psi^*_{i-1})$, then $f_{wc}(e^*_{i}\mid \psi'_b\cup\psi^*_{i-1})=0$ and for all $t\in[b]$, $f_{wc}(e^*_{i}\mid \psi'_{t-1})\geq 0$ due to $f$ is worst-case monotone, thus (\ref{eq:1}) also holds.

In  \citep{calinescu2007maximizing}, it has been shown that there exists a sequence of sets $M(e'_1), M(e'_2), \cdots, M(e'_b)$ whose nonempty members partition $S^*_{b^*}$,
such that for all $t\in[b]$ and $j \in[b^*]$ such that $e^*_{j}\in M(e'_t)$, we have (1) $\{e^*_{j}\}\cup S_{t-1} \in \mathcal{I}$, which implies that $\{e^*_{j}\}\cup \mathrm{dom}(\psi'_{t-1}) \in \mathcal{I}$ due to $S_{t-1}=\mathrm{dom}(\psi'_{t-1})$, and (2) $|M(e'_t)|\leq p$. Now consider a fixed $t\in[b]$ and any $e^*_{j} \in M(e'_t)$, because  $\{e^*_{j}\}\cup \mathrm{dom}(\psi'_{t-1}) \in \mathcal{I}$, $f_{wc}(e^*_{j}\mid \psi'_{t-1}) \leq   \max_{e\in E: \{e\}\cup \mathrm{dom}(\psi'_{t-1}) \in \mathcal{I}} f_{wc}(e\mid \psi'_{t-1})=  f_{wc}(e'_t\mid \psi'_{t-1})$ where the equality is due to the selection rule of $e'_t$. This together with (\ref{eq:1}), e.g., for all $t\in[b]$, $f_{wc}(e^*_{j}\mid \psi'_b\cup\psi^*_{j-1}) \leq  f_{wc}(e^*_{j}\mid \psi'_{t-1})$,  implies that
\begin{eqnarray}
\label{eq:good} f_{wc}(e^*_{j}\mid \psi'_b\cup\psi^*_{j-1})\leq   f_{wc}(e'_t\mid \psi'_{t-1})
\end{eqnarray}

Because $|M(e'_t)|\leq p$, we have $\sum_{e^*_{j}\in M(e'_t)} f_{wc}(e^*_{j}\mid \psi'_b\cup\psi^*_{j-1})\leq \sum_{e^*_{j}\in M(e'_t)}  f_{wc}(e'_t\mid \psi'_{t-1}) \leq p f_{wc}(e'_t\mid \psi'_{t-1})$ where the second inequality is due to (\ref{eq:good}). It follows that
\begin{eqnarray}
\label{eq:great}
\sum_{j\in [b^*]} f_{wc}(e^*_{j}\mid \psi'_b\cup\psi^*_{j-1}) = \sum_{t\in[b]}\sum_{e^*_{j}\in M(e'_t)} f_{wc}(e^*_{j}\mid \psi'_b\cup\psi^*_{j-1}) \leq p \sum_{t\in[b]}f_{wc}(e'_t\mid \psi'_{t-1}) \leq f_{wc}(\pi^w)
\end{eqnarray}
The equality is due to the nonempty members of $M(e'_1), M(e'_2), \cdots, M(e'_b)$  partition $S^*_{b^*}$, the first inequality is due to (\ref{eq:good}), and the second inequality is due to (\ref{eq:patch}).
This together with (\ref{eq:e}) implies that $ \frac{1}{p+1} f_{wc}(\pi^*_{wc}) \leq f_{wc}(\pi^w)$. $\Box$

\subsection{Improved Results for Cardinality Constraint}

\begin{algorithm}[hptb]
\caption{ Adaptive Worst-case Greedy Policy for Cardinality Constraint $\pi^g$}
\label{alg:LPP3}
\begin{algorithmic}[1]
\STATE $t=1; \psi_0=\emptyset$.
\WHILE {$t\leq k$}
\STATE select $e_t \in \argmax_{e\in E} f_{wc}(e\mid \psi_{t-1})$;
\STATE $\psi_{t}\leftarrow \psi_{t-1}\cup\{(e_t, \Phi(e_t))\}$;
\STATE $t\leftarrow t+1$;
\ENDWHILE
\end{algorithmic}
\end{algorithm}

In this section, we provide enhanced results for the following worst-case adaptive submodular maximization problem subject to a cardinality constraint $k$.
\[\max_{\pi}\{f_{wc}(\pi)\mid |E(\pi, \phi)|\leq k \mbox{ for all realizations } \phi\in U^+\}\]

Note that a single cardinality constraint is a $1$-system constraint. We show that the approximation ratio of $\pi^w$ (Algorithm \ref{alg:LPP2}) can be further improved to $1-1/e$ under a single cardinality constraint.
For ease of presentation, we provide a simplified version $\pi^g$ of $\pi^w$ (Algorithm \ref{alg:LPP2}) in Algorithm \ref{alg:LPP3}. Note that  $\pi^g$ follows the same greedy rule as described in $\pi^w$ to select $k$ items.

\begin{theorem}
\label{thm:2}
If the utility function $f: 2^E\times O^E\rightarrow \mathbb{R}_{\geq 0}$  is worst-case monotone, worst-case submodular with respect to $p(\phi)$ and it satisfies the property of minimal dependency, then $f_{wc}(\pi^g) \geq (1-1/e)f_{wc}(\pi^*_{wc})$ subject to cardinality constraints.
\end{theorem}
\emph{Proof:} Assume $\phi'$ is the worst-case realization of $\pi^g$, i.e., $\phi'=\argmin_{\phi} f(E(\pi^g, \phi), \phi)$. For each $t\in[k]$, let $\psi'_t$ denote the partial realization of the first $t$ items $S_t$ selected by $\pi^g$ conditioned on $\phi'$.
Given any optimal policy $\pi^*_{wc}$ and a partial realization $\psi'_t$ after running $\pi^g$ for $t$ rounds, we construct a realization $\phi^*$ as follows. First, we ensure that $\phi^*$ is consistent with $\psi'_t$ by setting $\phi^*(e)=\phi'(e)$ for each $e\in \mathrm{dom}(\psi'_t)$. Next, we complete the construction of $\phi^*$ by simulating the execution of $\pi^*_{wc}$. Starting with $i=1$ and let $\psi^*_0=\emptyset$, assume the optimal policy $\pi^*_{wc}$ picks $e^*_i$ as the $i$th item after observing $\psi^*_{i-1}$, we set the state $\phi^*(e^*_i)$ of $e^*_i$ to  $\argmin_{o\in O(e, \psi'_t\cup \psi^*_{i-1})}f(S_t\cup S^*_{i-1}\cup\{e^*_i\}, \psi'_t\cup \psi^*_{i-1}\cup\{(e^*_i, o)\}) - f(S_t\cup S^*_{i-1}, \psi'_t\cup \psi^*_i)$. After each round $i$, we update the observation using $\psi^*_{i}\leftarrow \psi^*_{i-1}\cup\{(e^*_i, \phi^*(e^*_i))\}$. This construction process continues until $\pi^*_{wc}$ does not select new items. Intuitively, in each round $i$ of $\pi^*_{wc}$, we choose a state for $e^*_i$ to minimize the marginal utility of $e^*_i$ on top of the partial realization $\psi'_t\cup \psi^*_{i-1}$. Note that there may exist multiple realizations which meet the above description, we choose an arbitrary one as $\phi^*$.

To prove this theorem, it suffices to show that for all $t\in[k]$,
\begin{eqnarray}f(S_{t}, \psi'_{t})-f(S_{t-1}, \psi'_{t-1})\geq  \frac{ f_{wc}(\pi^*_{wc})-f(S_{t-1}, \psi'_{t-1})}{k} \label{eq:f}\end{eqnarray} This is because by induction on $t$, we have that for any $l\in [k]$,
\begin{eqnarray}
\label{eq:last}
\sum_{t\in [l]} \{ f(S_{t}, \psi'_{t})-f(S_{t-1}, \psi'_{t-1})\}\geq (1-e^{-l/k}) f_{wc}(\pi^*_{wc})
\end{eqnarray}
Then this theorem follows because $f_{wc}(\pi^g)=f(E(\pi^g, \phi'), \phi') = f(S_k, \psi'_k) = \sum_{t\in [k]} \{ f(S_{t}, \psi'_{t})-f(S_{t-1}, \psi'_{t-1})\}\geq (1-1/e) f_{wc}(\pi^*_{wc})$.

Thus, we focus on proving (\ref{eq:f}) in the rest of the proof.
\begin{eqnarray}
&&f(S_{t}, \psi'_{t})-f(S_{t-1}, \psi'_{t-1})=  f(S_{t-1}\cup \{e'_{t}\}, \psi'_{t-1}\cup\{(e'_{t}, \phi'(e'_{t}))\})-f(S_{t-1}, \psi'_{t-1})~\nonumber\\
&&\geq \min_{o\in O(e'_t, \psi'_{t-1})}\{f(S_{t-1}\cup \{e'_{t}\}, \psi'_{t-1}\cup\{(e'_{t}, o)\})-f(S_{t-1}, \psi'_{t-1})\}~\nonumber\\
&&= \max_{e\in E}f_{wc}(e\mid \psi'_{t-1})\geq \frac{\sum_{i\in [k]}f_{wc}(e^*_i\mid \psi'_{t-1})}{k}\geq \frac{\sum_{i\in [k]}f_{wc}(e^*_i\mid \psi'_{t-1}\cup\psi^*_{i-1})}{k}~\nonumber\\
&&=\frac{ f(S_{t-1}\cup S^*_k, \psi'_{t-1}\cup\psi^*_k)-f(S_{t-1}, \psi'_{t-1})}{k}\geq \frac{ f( S^*_k, \psi^*_k)-f(S_{t-1}, \psi'_{t-1})}{k} \geq  \frac{ f_{wc}(\pi^*_{wc})-f(S_{t-1}, \psi'_{t-1})}{k}~\nonumber
\end{eqnarray}
The second equality is due to the selection rule of $e'_t$, i.e., $e'_t \in \argmax_{e\in E} f_{wc}(e\mid \psi'_{t-1})$ and the fourth inequality is due to $f: 2^E\times O^E\rightarrow \mathbb{R}_{\geq 0}$  is worst-case monotone. $\Box$

\paragraph{A faster algorithm that maximizes the expected worst-case utility.} Inspired by the sampling technique developed in \citep{mirzasoleiman2015lazier,tang2021beyond}, we next present a faster randomized algorithm that achieves a  $1-1/e-\epsilon$ approximation ratio when maximizing the \emph{expected} worst-case utility subject to a cardinality constraint. We first extend the definition of $\pi$ so that it can represent any randomized policies. In particular, we re-define $\pi$ as a mapping function that  maps a set of partial realizations  to a distribution $\mathcal{P}(E)$ of $E$: $\pi: 2^{E}\times O^E \rightarrow \mathcal{P}(E)$. The expected worst-case  utility $\tilde{f}_{wc}(\pi)$ of a randomized policy $\pi$  is defined as
\begin{eqnarray}
\tilde{f}_{wc}(\pi)=\min_{\phi\in U^+}\mathbb{E}_{\Pi}[f(E(\pi, \phi), \phi)]~\nonumber
\end{eqnarray}
where the expectation is taken over the internal randomness of the policy $\pi$.

\begin{algorithm}[hptb]
\caption{Adaptive Stochastic Worst-case Greedy Policy $\pi^{f}$}
\label{alg:LPP13}
\begin{algorithmic}[1]
\STATE $t=1; \psi_0=\emptyset$.
\WHILE {$t \leq k$}
\STATE $H \leftarrow$ a random set of size $\frac{n}{k}\log\frac{1}{\epsilon}$ sampled uniformly at random  from $E$;
\STATE select $e_t \in \argmax_{e\in H} f_{wc}(e\mid \psi_{t-1})$;
\STATE $\psi_{t}\leftarrow \psi_{t-1}\cup\{(e_t, \Phi(e_t))\}$;
\STATE $t\leftarrow t+1$;
\ENDWHILE
\end{algorithmic}
\end{algorithm}

Now we are ready to present our \emph{Adaptive Stochastic Worst-case Greedy Policy} $\pi^{f}$. Starting with an empty set and  at each round $t\in [k]$,  $\pi^{f}$ first samples a random set $H$ of size $\frac{n}{k}\log\frac{1}{\epsilon}$  uniformly at random from $E$, where $\epsilon>0$ is some positive constant. Then it selects the $t$-th item $e_t$ from $H$ that maximizes the worst-case marginal utility  on top of the current observation $\psi_{t-1}$, i.e., $e_t \in \argmax_{e\in H} f_{wc}(e\mid \psi_{t-1})$. This process continues until  $\pi^{f}$ selects $k$ items.

We next show that the expected worst-case utility of $\pi^f$ is at least $(1-1/e-\epsilon)f_{wc}(\pi^*_{wc})$. Moreover, the running time of $\pi^f$ is bounded by $O(n\log\frac{1}{\epsilon})$ which is \emph{linear} in the problem size $n$  and \emph{independent} of the cardinality constraint $k$. We move the proof of the following theorem to appendix.
\begin{theorem}
\label{thm:important}
For cardinality constraints, if the utility function $f: 2^E\times O^E\rightarrow \mathbb{R}_{\geq 0}$  is worst-case monotone, worst-case submodular with respect to $p(\phi)$ and it satisfies the property of minimal dependency, then $\tilde{f}_{wc}(\pi^f) \geq (1-1/e-\epsilon)f_{wc}(\pi^*_{wc})$. The running time of $\pi^f$ is bounded by $O(n\log\frac{1}{\epsilon})$.
\end{theorem}

\subsection{Pointwise submodularity does not imply worst-case submodularity}
\label{sec:doesnot}
In this section, we first introduce the notations of \emph{pointwise submodularity} and \emph{pointwise monotonicity}. Then we construct an example to show that the performance of $\pi^g$ could be arbitrarily bad even if $f: 2^E\times O^E\rightarrow \mathbb{R}_{\geq 0}$ is worst-case monotone, pointwise submodular and it satisfies the property of minimal dependency.  As a result, the main result claimed in \citep{cuong2014near} does not hold in general. This observation, together with Theorem \ref{thm:2},  also indicates that pointwise submodularity does not imply worst-case submodularity. In Section \ref{sec:app}, we show that adding one more condition, i.e., for all $e\in E$ and all possible partial realizations $\psi$ such that $e\notin \mathrm{dom}(\psi)$, $ O(e, \psi)=O(e, \emptyset)$, on top of the above three conditions is sufficient to ensure the worst-case submodularity of $f: 2^E\times O^E\rightarrow \mathbb{R}_{\geq 0}$.
\begin{definition}\citep{golovin2011adaptive}\emph{[Pointwise Submodularity and Pointwise Monotonicity]}
\label{def:121}
A function $f: 2^E\times O^E\rightarrow \mathbb{R}_{\geq 0}$  is  pointwise submodular if $f(S, \phi)$ is submodular for any given realization $\phi\in U^+$. Formally, consider any realization $\phi\in U^+$, any two sets $E_1\subseteq E$ and $E_2 \subseteq E$ such that $E_1 \subseteq E_2$, and any item $e\notin E_2$, we have $f(E_1\cup\{e\}, \phi)- f(E_1, \phi) \geq f(E_2\cup\{e\}, \phi)- f(E_2, \phi)$. We say $f: 2^E\times O^E\rightarrow \mathbb{R}_{\geq 0}$ is pointwise monotone if for all $E_1\subseteq E$, $e\notin E_1$ and $\phi$ such that $p(\phi)>0$, $f(E_1\cup\{e\}, \phi)\geq f(E_1, \phi)$.
\end{definition}

\textbf{Example.} Assume the ground set is composed of three items $E=\{e_1, e_2, e_3\}$, there are two possible states $O=(o_1, o_2)$ and there are three realizations with positive probability $U^+=\{\{(e_1, o_1), (e_2, o_2), (e_3, o_2)\},$
$ \{(e_1, o_2), (e_2, o_1), (e_3, o_2)\}, $
$\{(e_1, o_2), (e_2, o_2), (e_3, o_1)\}\}$. The utility function $f: 2^E\times O^E\rightarrow \mathbb{R}_{\geq 0}$ is defined in Table \ref{rrr}  where $\epsilon$ is some positive constant that is smaller than $1$.
\begin{table*}
\begin{center}
\begin{tabular}{ |c|c|c|c|}
\hline
 & $\{(e_1, o_1), (e_2, o_2), (e_3, o_2)\}$ & $\{(e_1, o_2), (e_2, o_1), (e_3, o_2)\}$ & $\{(e_1, o_2), (e_2, o_2), (e_3, o_1)\}$\\
\hline
$\emptyset$ & $0$ & $0$ & $0$\\
\hline
$\{e_1\}$& $\epsilon$& $\epsilon$& $\epsilon$ \\
\hline
$\{e_2\}$& $1$ & $0$ & $1$ \\
\hline
$\{e_3\}$ & $1$ & $1$ & $0$\\
\hline
$\{e_1, e_2\}$ & $1+\epsilon$ & $\epsilon+0$ & $1+\epsilon$\\
\hline
$\{e_2, e_3\}$ & $1+1$ & $1+0$ & $1+0$\\
\hline
$\{e_1, e_3\}$ & $1+\epsilon$ & $1+\epsilon$ & $\epsilon+0$\\
\hline
$\{e_1, e_2, e_3\}$ & $1+1+\epsilon$ & $1+\epsilon$ & $1+\epsilon$\\
\hline
\end{tabular}
\caption{The definition of $f: 2^E\times O^E\rightarrow \mathbb{R}_{\geq 0}$.}
\label{rrr}
\end{center}
\end{table*}

Note that in the above example, $f(S, \phi)$ is a linear function for any given realization $\phi\in U^+$, thus it is pointwise submodular. Moreover, it is easy to verify that this function is worst-case monotone and it satisfies the property of minimal dependency. Assume the cardinality constraint is $k=2$. According to the design of $\pi^g$, it first picks $e_1$. This is because the worst-case marginal utility of $e_1$ on top of $\emptyset$ is $f_{wc}(e_1\mid \emptyset )=\epsilon$, which is larger than $f_{wc}(e_2\mid \emptyset )=0$ and $f_{wc}(e_3\mid \emptyset )=0$. Then it picks $e_2$ as the second item regardless of the realization of $\Phi(e_1)$, this is because $f_{wc}(e_2\mid \{(e_1, o_1)\} )=f_{wc}(e_3\mid \{(e_1, o_1)\} )=1$ and $f_{wc}(e_2\mid \{(e_1, o_2)\} )=f_{wc}(e_3\mid \{(e_1, o_2)\} )=0$, i.e., $e_2$ and $e_3$ have the same worst-case marginal utility on top of $\{(e_1, o_1)\}$ or $\{(e_1, o_2)\}$. Thus, $f_{wc}(\pi^g)=f(\{e_1, e_2\}, \{(e_1, o_2), (e_2, o_1), (e_3, o_2)\})=\epsilon$. However, the optimal solution $\pi^*_{wc}$ always picks $\{e_2, e_3\}$ which has the worst-case utility $f_{wc}(\pi^*_{wc})=f(\{e_2, e_3\}, \{(e_1, o_2), (e_2, o_1), (e_3, o_2)\})=1$. Thus, the approximation ratio of $\pi^g$ is $\frac{\epsilon}{1}$ whose value is arbitrarily close to $0$ as $\epsilon$ tends to $0$.

\subsection{Applications}
\label{sec:app}
We next discuss several important applications whose utility function satisfies all three conditions stated in Theorem \ref{thm:1}, Theorem \ref{thm:2} and Theorem \ref{thm:important}, i.e.,  worst-case monotonicity, worst-case submodularity and minimal dependency. All proofs are moved to appendix.
\subsubsection{Pool-based Active Learning.}
 \label{sec:activelearning}We consider a set $\mathcal{H}$ of candidate hypothesis. Let $E$ denote a set of data points and each data point $e\in E$ has a random label $\Phi(e)$. Each
hypothesis $h \in \mathcal{H}$ represents some realization, i.e., $h : E \rightarrow \Phi$. We use $p_H$ to denote a prior distribution over
hypothesis. Let $p_H(\mathcal{H}') = \sum_{h\in \mathcal{H}'} p_H(h)$
for any $\mathcal{H}' \subseteq \mathcal{H}$. Let $p(\phi) = p_H(h| \phi\sim h)$ denote the prior distribution over realizations. Define the version space $\mathcal{H}(\psi)$ under observations $\psi$ as a set that contains all hypothesis whose labels are consistent with $\psi$ in the domain of $\psi$, i.e., $\mathcal{H}(\psi) = \{h\in \mathcal{H}| h\sim \psi\}$. We next define  the utility function $f(S, \phi)$ of generalized binary search under the Bayesian setting for a given realization $\phi$ and a group of labeled data points $S$:
\begin{eqnarray}
f(S, \phi) =  1- p_H(\mathcal{H}(\phi(S))) \label{eq:10}
\end{eqnarray}
The above utility function measures the reduction in version space mass after obtaining the states (a.k.a. labels) about $S$ conditioned on a realization $\phi$. Our goal is to sequentially query the labels of at some data points, each selection is based past feedback, to maximize the reduction in version
space mass.
\begin{proposition}
\label{pro:1}
The utility function $f: 2^{E}\times O^E\rightarrow \mathbb{R}_{\geq0}$ of pool-based active learning  is worst-case submodular with respect to $p(\phi)$.
\end{proposition}

 \subsubsection{The Case when $O(e, \psi)=O(e, \emptyset)$.} We next show that the utility function $f: 2^{E}\times O^E\rightarrow \mathbb{R}_{\geq0}$  is worst-case submodular with respect to $p(\phi)$ if the following three conditions are satisfied:
 \begin{enumerate}
  \item  For all $e\in E$ and all possible partial realizations $\psi$ such that $e\notin \mathrm{dom}(\psi)$, $ O(e, \psi)=O(e, \emptyset)$.
   \item $f: 2^{E}\times O^E\rightarrow \mathbb{R}_{\geq0}$ is pointwise submodular and pointwise monotone with respect to $p(\phi)$.
   \item $f: 2^{E}\times O^E\rightarrow \mathbb{R}_{\geq0}$ satisfies the property of minimal dependency.
\end{enumerate}

\begin{proposition}
\label{pro:2}
If the above three conditions are satisfied, then $f: 2^{E}\times O^E\rightarrow \mathbb{R}_{\geq0}$ is worst-case monotone and worst-case submodular  with respect to $p(\phi)$ and it satisfies the property of minimal dependency.
\end{proposition}

 It is easy to verify that the first condition is satisfied if the states of all items are independent of each other. This indicates that the utility function of the classic Stochastic Submodular Maximization problem \citep{asadpour2016maximizing} satisfies the properties of worst-case monotonicity and worst-case submodularity. We next present two applications of the Stochastic Submodular Maximization problem.

\paragraph{The Sensor Selection Problem with Unreliable Sensors \citep{golovin2011adaptive}.} In this application, we would like to monitor a spatial phenomenon such as humidity in a house by selecting a set of $k$ most ``informative'' sensors from a ground set $E$. We assume that sensors are unreliable and they may fail to work, but we must deploy a sensor before finding out whether it has failed or not. We assign a state $\Phi(e)$ to each sensor $e$ to represent the status of $e$. For example, $\Phi(e)$ may take values from $\{\textsf{working}, \textsf{failed}\}$. 
Suppose that each sensor has a known probability of failure and the status of all sensors are independent of each other, our goal is to sequentially deploy a group of $k$ sensors to maximize the informativeness of functioning sensors. Here, we quantify the informativeness of a set  of working sensors using a monotone and submodular set function.

\paragraph{Stochastic Maximum $k$-Cover \citep{asadpour2008stochastic}.} The input of the classic maximum $k$-cover problem is a collection $E$ of subsets of $\{1, 2, \cdots, n\}$, our goal is to find $k$ subsets from $E$ so as to maximize the size of their union \citep{feige1998threshold}. Under the stochastic setting, the subset that an element of  $E$ can cover is uncertain, one must select an element before revealing the actual subset that can be covered by that element. Our goal is to sequentially select a group of $k$ elements  from $E$ such that the expected size of their union is maximized.

\paragraph{Adaptive Match-Making \citep{golovin2011adaptive1}.} In this application, we consider an adaptive match-making problem such as online dating. The input of this problem is an undirected graph and for any pair of nodes $u$ and $v$, there is an edge $(u,v)$ if and only if $v$ meets the
requirements specified by $u$ and vice-versa. We assign a state $\Phi((u,v))$ to each edge $(u,v)$ to represent the matching score of $(u,v)$. The service will sequentially select a set of edges (corresponding to matches), and after each
selection, we can observe the actual matching score through based on the feedback from the participants. Each node can specify maximum number of dates s/he is willing to receive. Suppose the states of all edges are independent of each other, our goal is to maximize the expected  value of the sum of the matching scores of all selected edges.

 \subsubsection{Adaptive Viral Marketing.} In this application, our goal is to select a group of $k$ influential users from a social network to help promote some product or information. Let $G=(E, Z)$ denote a social network, where $E$  is a set of nodes and $Z$ is a set of edges.  We use Independent Cascade  Model (IC)  \citep{kempe2008cascade} to model the diffusion process in a social network. Under IC model, we assign a propagation probability $p(u,v)\in[0,1]$ to each edge $(u,v)\in Z$ such that $(u,v)$ is ``live'' with probability $p(u,v)$ and $(u,v)$ is ``blocked'' with probability $1-p(u,v)$. The state $\phi(u)$ of a node $u$ is a function $\phi_u: Z\rightarrow \{0, 1, ?\}$ such that $\phi_u((u,v))=0$ indicates that $(u,v)$ is blocked (i.e., $u$ fails to influence $v$), $\phi_u((u,v))=1$ indicates that $(u,v)$ is live (i.e., $u$ succeeds in influencing $v$), and $\phi_u((u,v))=?$ indicates that  selecting $u$ can not reveal the status of $(u,v)$. Consider a set $S$ of nodes which are selected initially and a realization $\phi$,  the utility $f(S, \phi)$ represents the number of individuals that can be reached by at least one node from $S$ through live edges, i.e.,
\begin{eqnarray}
\label{eq:11}
  f(S, \phi)=|\{v| \exists u\in S, w\in E, \phi_u((w, v))=1 \}|+|S|
\end{eqnarray}

 \begin{proposition}
\label{pro:3}
 The utility function $f: 2^{E}\times O^E\rightarrow \mathbb{R}_{\geq0}$ of adaptive viral marketing is worst-case monotone and worst-case submodular  with respect to $p(\phi)$ and it satisfies the property of minimal dependency.
\end{proposition}

\subsection{Hardness of Approximation}
In this section, we discuss the hardness results for the general worst-case adaptive submodular maximization problem. We first discuss the case of $p$-system constraints. \cite{badanidiyuru2014fast} gave a simple proof that the factor of $1/p$ is optimal for maximizing a monotone and submodular function subject to a  $p$-system constraint. This upper bound also applies to our setting  because 
one can consider the traditional non-adaptive optimization problem as a special case of the  adaptive optimization problem such that the distributions of realizations are deterministic.  Note that if the distributions are deterministic, then worst-case scenario coincides  with average-case scenario, and in this special case, we can show that worst-case submodularity and worst-case monotonicity reduce to the classical notion of monotone submodular set functions.  Formally, given any monotone and submodular function $\hat{f}: 2^{E}\rightarrow \mathbb{R}_{\geq0}$, we can construct an equivalent function $f: 2^{E}\times O^E\rightarrow \mathbb{R}_{\geq0}$ such that there is only one realization $\phi$, and for any subset of items $S\subseteq E$, define $f(S, \phi)=\hat{f}(S)$. Consider any two partial realizations $\psi$ and $\psi'$ such that $\psi\subseteq \psi'$. For each $e\in E\setminus \mathrm{dom}(\psi')$, we have
\begin{eqnarray*}
f_{wc}(e\mid \psi) &=& \hat{f}(\{e\}\cup \mathrm{dom}(\psi))-\hat{f}(\mathrm{dom}(\psi)) \\
&\geq&  \hat{f}(\{e\}\cup \mathrm{dom}(\psi'))-\hat{f}(\mathrm{dom}(\psi')) \\
&=& f_{wc}(e\mid \psi')
\end{eqnarray*}
where the inequality is due to the assumption that $\hat{f}: 2^{E}\rightarrow \mathbb{R}_{\geq0}$ is submodular. Moreover, $f_{wc}(e\mid \psi) = \hat{f}(\{e\}\cup \mathrm{dom}(\psi))-\hat{f}(\mathrm{dom}(\psi)) \geq 0$ due to $\hat{f}: 2^{E}\rightarrow \mathbb{R}_{\geq0}$ is monotone. Hence, $f: 2^{E}\times O^E\rightarrow \mathbb{R}_{\geq0}$ is worst-case monotone and worst-case submodular. Given the upper bound of $1/p$, our worst-case greedy policy $\pi^w$, whose approximation ratio is $1/(p+1)$, achieves nearly optimal performance.

We next discuss the case of cardinality constraints. It is well known that the factor of $1-1/e$ is optimal for maximizing a monotone and submodular function subject to a  cardinality constraint \citep{feige1998threshold}. Again, because the traditional non-adaptive optimization problem is a special case of our problem, this upper bound also holds under our setting. This indicates that our greedy policy $\pi^g$, which approximates the optimum to within a factor of $1-1/e$, is optimal.

\section{Robust Adaptive Submodular Maximization}
\label{sec:4}
\subsection{Cardinality Constraint}
\begin{algorithm}[hptb]
\caption{ Adaptive Average-case Greedy Policy $\pi^a$ \citep{golovin2011adaptive}}
\label{alg:LPP4}
\begin{algorithmic}[1]
\STATE $t=1; \psi_0=\emptyset$.
\WHILE {$t\leq \lceil\frac{k}{2}\rceil$}
\STATE select $e_t \in \argmax_{e\in E} f_{avg}(e\mid \psi_{t-1})$;
\STATE $\psi_{t}\leftarrow \psi_{t-1}\cup\{(e_t, \Phi(e_t))\}$;
\STATE $t\leftarrow t+1$;
\ENDWHILE
\end{algorithmic}
\end{algorithm}
\begin{algorithm}[hptb]
\caption{ Adaptive Hybrid Policy for Cardinality Constraint $\pi^h$}
\label{alg:LPP51}
\begin{algorithmic}[1]
\STATE Run $\pi^g$ to select  $\lfloor\frac{k}{2}\rfloor$ items.
\STATE Run $\pi^a$ to select  $\lceil\frac{k}{2}\rceil$ items, ignoring the observation obtained from running $\pi^g$.
\end{algorithmic}
\end{algorithm}
In this section, we study the robust adaptive submodular maximization problem subject to a cardinality constraint $k$, i.e.,
\[\max_{\pi}\{\alpha(\pi)\mid |E(\pi, \phi)|\leq k \mbox{ for all } \phi\in U^+\}\]

 Before presenting our solution, we first introduce an adaptive greedy policy $\pi^a$ from \citep{golovin2011adaptive}. A detailed description of $\pi^a$ can be found in Algorithm \ref{alg:LPP4}. $\pi^a$ selects $k$ items iteratively, in each round $t\in [k]$, it selects an item $e_t$ that maximizes the average-case marginal utility on top of the current partial realization $\psi_{t-1}$: $e_t \in \argmax_{e\in E} f_{avg}(e\mid \psi_{t-1})$. \cite{golovin2011adaptive} show that if $f: 2^{E}\times O^E\rightarrow \mathbb{R}_{\geq0}$ is adaptive monotone and adaptive submodular, then $\pi^a$ achieves a $1-e^{-\frac{\lceil\frac{k}{2}\rceil}{k}}$-approximation ratio, i.e.,
\begin{eqnarray}
\label{eq:lastlast}
 f_{avg}(\pi^a)\geq  (1-e^{-\frac{\lceil\frac{k}{2}\rceil}{k}})f_{avg}(\pi^*_{avg})
 \end{eqnarray}

Our hybrid policy $\pi^h$ (Algorithm \ref{alg:LPP51}) runs the adaptive worst-case greedy policy $\pi^g$ (Algorithm \ref{alg:LPP3}) to select $\lfloor\frac{k}{2}\rfloor$ items, then runs the adaptive average-case greedy policy $\pi^a$ (Algorithm \ref{alg:LPP4}) to select $\lceil\frac{k}{2}\rceil$ items, ignoring the observation obtained from running $\pi^g$. Thus, $\pi^h$ can also be represented as $\pi^g_{\lfloor\frac{k}{2}\rfloor}@\pi^a$ where $\pi^g_{\lfloor\frac{k}{2}\rfloor}$ denotes the level-$\lfloor\frac{k}{2}\rfloor$-truncation of $\pi^g$. We next show that the robustness ratio $\alpha(\pi^h)$ of $\pi^h$ is at least $1-e^{-\frac{\lfloor\frac{k}{2}\rfloor}{k}}$ whose value approaches $1-e^{-\frac{1}{2}}$ when $k$ is large.
\begin{theorem}
\label{thm:4}
If $f: 2^{E}\times O^E\rightarrow \mathbb{R}_{\geq0}$ is worst-case monotone, worst-case submodular and adaptive submodular with respect to $p(\phi)$, and it satisfies the property of minimal dependency,  then  $\alpha(\pi^h)\geq 1-e^{-\frac{\lfloor\frac{k}{2}\rfloor}{k}}$ subject to a cardinality constraint $k$.
\end{theorem}
\emph{Proof:} Assume $\phi'$ is the worst-case realization of $\pi^h$, i.e., $\phi'=\argmin_{\phi} f(E(\pi^h, \phi), \phi)$. Then because $f: 2^{E}\times O^E\rightarrow \mathbb{R}_{\geq0}$ is worst-case monotone and worst-case submodular with respect to $p(\phi)$ and it satisfies the property of minimal dependency, we have
\begin{eqnarray}
\label{eq:end}
f_{wc}(\pi^h)&& = f(E(\pi^h, \phi'), \phi') \geq f(E(\pi^g_{\lfloor\frac{k}{2}\rfloor}, \phi'), \phi')\geq (1-e^{-\frac{\lfloor\frac{k}{2}\rfloor}{k}})f_{wc}(\pi^*_{wc})
\end{eqnarray}
The first inequality is due to $\pi^h$ can be represented as $\pi^g_{\lfloor\frac{k}{2}\rfloor}@\pi^a$ and $f: 2^{E}\times O^E\rightarrow \mathbb{R}_{\geq0}$ is worst-case monotone with respect to $p(\phi)$. The second inequality is due to (\ref{eq:last}).
Moreover, because worst-case monotonicity implies adaptive monotonicity (see Section \ref{sec:3.2}), if $f: 2^{E}\times O^E\rightarrow \mathbb{R}_{\geq0}$ is worst-case monotone and adaptive submodular with respect to $p(\phi)$, $f: 2^{E}\times O^E\rightarrow \mathbb{R}_{\geq0}$ is adaptive monotone and adaptive submodular with respect to $p(\phi)$. Then if $f: 2^{E}\times O^E\rightarrow \mathbb{R}_{\geq0}$ is worst-case monotone and adaptive submodular with respect to $p(\phi)$, we have
\begin{eqnarray}
\label{eq:end2}
f_{avg}(\pi^h) \geq f_{avg}(\pi^a) \geq (1-e^{-\frac{\lceil\frac{k}{2}\rceil}{k}})f_{avg}(\pi^*_{avg})
\end{eqnarray}
The first inequality is due to $\pi^h$ can be represented as $\pi^g_{\lfloor\frac{k}{2}\rfloor}@\pi^a$ and $f: 2^{E}\times O^E\rightarrow \mathbb{R}_{\geq0}$ is worst-case monotone with respect to $p(\phi)$. The second inequality is due to (\ref{eq:lastlast}).

(\ref{eq:end}) and (\ref{eq:end2}) imply that $\alpha(\pi^h)=\min\{\frac{f_{wc}(\pi^h)}{f_{wc}(\pi^*_{wc})}, \frac{f_{avg}(\pi^h)}{f_{avg}(\pi^*_{avg})}\} =\min\{1-e^{-\frac{\lfloor\frac{k}{2}\rfloor}{k}}, 1-e^{-\frac{\lceil\frac{k}{2}\rceil}{k}}\}\geq 1-e^{-\frac{\lfloor\frac{k}{2}\rfloor}{k}}$.
 $\Box$

\subsection{Partition Matroid Constraint}
\begin{algorithm}[hptb]
\caption{ Adaptive Average-case Greedy Policy for Partition Matroid $\pi^{ma}$}
\label{alg:LPP5}
\begin{algorithmic}[1]
\STATE $t=1; z=1; \psi_{1, 0}=\emptyset$.
\WHILE {$z \leq b$}
\WHILE {$t \leq \lceil\frac{k_z}{2}\rceil$}
\STATE select $e_{z,t} \in \argmax_{e\in E_z} f_{avg}(e\mid \psi_{z, t-1})$;
\STATE $\psi_{z, t}\leftarrow \psi_{z, t-1} \cup\{(e_{z,t}, \Phi(e_{z,t}))\}$;
\STATE $t\leftarrow t+1$;
\ENDWHILE
\STATE $z\leftarrow z+1$;
\ENDWHILE
\end{algorithmic}
\end{algorithm}

\begin{algorithm}[hptb]
\caption{ Adaptive Worst-case Greedy Policy for Partition Matroid $\pi^{mw}$}
\label{alg:LPP7}
\begin{algorithmic}[1]
\STATE $t=1; z=1; \psi_{1, 0}=\emptyset$.
\WHILE {$z \leq b$}
\WHILE {$t \leq \lfloor\frac{k_z}{2}\rfloor$}
\STATE select $e_{z,t} \in \argmax_{e\in E_z} f_{wc}(e\mid \psi_{z, t-1})$;
\STATE $\psi_{z, t}\leftarrow \psi_{z, t-1} \cup\{(e_{z,t}, \Phi(e_{z,t}))\}$;
\STATE $t\leftarrow t+1$;
\ENDWHILE
\STATE $z\leftarrow z+1$;
\ENDWHILE
\end{algorithmic}
\end{algorithm}

\begin{algorithm}[hptb]
\caption{Adaptive Hybrid Policy for Partition Matroid $\pi^m$}
\label{alg:LPP6}
\begin{algorithmic}[1]
\STATE Run $\pi^{mw}$;
\STATE Run $\pi^{ma}$, ignoring the observation obtained from running $\pi^{mw}$;
\end{algorithmic}
\end{algorithm}

We next study the robust adaptive submodular maximization problem subject to a partition matroid constraint $(E, \mathcal{I})$. Let $E_1, E_2, \ldots, E_b$ be a collection of disjoint subsets of $E$. Given a set of $b$ integers $\{k_z\mid z\in [b]\}$, let $ \mathcal{I}$ be a collection of subsets of $E$ such that for each $I\in \mathcal{I}$, we  have $\forall z\in [b]$, $|E_z \cap I| \leq k_z$.  Our problem  can be formulated as follows:
 \[\max_{\pi}\{\alpha(\pi)\mid E(\pi, \phi)\in \mathcal{I} \mbox{ for all } \phi\in U^+\}\]

 To solve this problem, we first introduce two subroutines. Our final policy is a concatenation  of these two subroutines.

 The first subroutine is an adaptive average-case greedy policy $\pi^{ma}$.  $\pi^{ma}$ runs in $b$ meta-rounds, and in each meta-round $z\in[b]$, it selects  $\lceil\frac{k_z}{2}\rceil$ items from $E_z$ in   $\lceil\frac{k_z}{2}\rceil$ iterations such that in each iteration $t\in [\lceil\frac{k_z}{2}\rceil]$ it selects an item $e_{z,t}$ that maximizes the average-case marginal utility on top of the current partial realization $\psi_{z, t-1}$: $e_{z,t} \in \argmax_{e\in E_z} f_{avg}(e\mid \psi_{z, t-1})$. A detailed description of $\pi^{ma}$ can be found in Algorithm \ref{alg:LPP5}.

 The second subroutine is an adaptive worst-case greedy policy $\pi^{mw}$.  $\pi^{mw}$ runs in $b$ meta-rounds, and in each meta-round $z\in[b]$, it selects  $\lfloor\frac{k_z}{2}\rfloor$ items from $E_z$ in   $\lfloor\frac{k_z}{2}\rfloor$ iterations such that in each iteration $t\in [\lfloor\frac{k_z}{2}\rfloor]$ it selects an item $e_{z,t}$ that maximizes the worst-case marginal utility on top of the current partial realization $\psi_{z, t-1}$: $e_{z,t} \in \argmax_{e\in E_z} f_{wc}(e\mid \psi_{z, t-1})$.  A detailed description of $\pi^{mw}$ can be found in Algorithm \ref{alg:LPP7}.

 Our hybrid policy $\pi^m$ (Algorithm \ref{alg:LPP6}) runs $\pi^{mw}$ first, then runs  $\pi^{ma}$, ignoring the observation obtained from running $\pi^{mw}$. Thus, $\pi^m$ can also be represented as $\pi^{mw}@\pi^{ma}$. We next present the main result of this section. 
\begin{theorem}
\label{thm:5}
Let $\gamma=\min_{z\in[b]} \frac{\lfloor\frac{k_z}{2}\rfloor}{k_z}$. If $f: 2^{E}\times O^E\rightarrow \mathbb{R}_{\geq0}$ is worst-case monotone, worst-case submodular and adaptive submodular with respect to $p(\phi)$, and it satisfies the property of minimal dependency,  then  $\alpha(\pi^m)\geq \frac{\gamma}{\gamma+1}$ subject to partition matroid constraints, where  $\frac{\gamma}{\gamma+1}$ approaches $1/3$  when $\min_{z\in[b]} k_z$ is large,.
\end{theorem}

To prove this theorem, it suffices to show that $f_{wc}(\pi^*_{wc}) \leq  (1+\frac{1}{\gamma}) f_{wc}(\pi^m)$ and $f_{avg}(\pi^*_{avg}) \leq  3  f_{avg}(\pi^m)$. This is because if these two results hold, then $\alpha(\pi^m)=\min\{\frac{f_{wc}(\pi^m)}{f_{wc}(\pi^*_{wc})}, \frac{f_{avg}(\pi^m)}{f_{avg}(\pi^*_{avg})}\} \geq \min\{\gamma/(1+\gamma), 1/3\}\geq \gamma/(1+\gamma)$. The rest of this section is devoted to proving these two results.

\begin{lemma}
\label{lem:vijya1}
If $f: 2^{E}\times O^E\rightarrow \mathbb{R}_{\geq0}$ is worst-case monotone and worst-case submodular with respect to $p(\phi)$, and it satisfies the property of minimal dependency,  then $f_{wc}(\pi^*_{wc}) \leq  (1+\frac{1}{\gamma}) f_{wc}(\pi^m)$. \end{lemma}
\emph{Proof:} Because $f: 2^{E}\times O^E\rightarrow \mathbb{R}_{\geq0}$ is worst-case monotone, we have $f_{wc}(\pi^m)=f_{wc}(\pi^{mw}@\pi^{ma})\geq f_{wc}(\pi^{mw})$. To prove this lemma, it suffices to show that $f_{wc}(\pi^*_{wc}) \leq  (1+\frac{1}{\gamma}) f_{wc}(\pi^{mw})$. We next focus on proving this result.

Assume $\phi'$ is the worst-case realization of $\pi^{mw}$, i.e., $\phi'=\argmin_{\phi} f(E(\pi^{mw}, \phi), \phi)$. For each $t\in[k]$, let $e_{z,t}$ denote the $t$-th item selected by $\pi^{mw}$ in meta-round $z$ conditioned on $\phi'$, and  let $\psi'_{z,t}$ denote the partial realization of all items selected by $\pi^{mw}$ before (including) the $t$-th iteration in meta-round $z$. Moreover, let $\psi'$ denote the final realization after running $\pi^{mw}$ conditioned on $\phi'$. 
Given  $\psi'$ and the optimal worst-case solution $\pi^*_{wc}$, we build a realization $\phi^*$ based on $\psi'$ according to the same procedures described in the proof of Theorem \ref{thm:1}. Let $\{S_1^*, S_2^*,\cdots, S_b^*\}$ denote the set of items selected by $\pi^*_{wc}$ conditioned on $\phi^*$, where for each $z\in[b]$,  $S_z^*$ represents the set of items selected from $E_z$. 

Hence, for each $z\in[b]$ and any $e^*_l\in S^*_{z}$, and for any $t\in[|\lfloor\frac{k_z}{2}\rfloor|]$, we have
 {\small\begin{eqnarray}
 \label{eq:kdd}
  f_{wc}(e^*_l \mid \psi^*_{l-1} \cup \psi') \leq f_{wc}(e^*_l \mid \psi'_{z, |\lfloor\frac{k_z}{2}\rfloor|}) \leq  f_{wc}(e^*_l \mid  \psi'_{z, t-1}) \leq   \max_{e'\in E_z} f_{wc}(e'\mid \psi'_{z, t-1})=  f_{wc}(e_{z,t}\mid \psi'_{z, t-1})
  \end{eqnarray}}
  where $\psi^*_{l-1}$ is the partial realization before $\pi^*_{wc}$ selects $e^*_l$. The first inequality is due to inequality is due to  $f: 2^{E}\times O^E\rightarrow \mathbb{R}_{\geq0}$ is  worst-case submodular with respect to $p(\phi)$ and $\psi'_{z, |\lfloor\frac{k_z}{2}\rfloor|} \subseteq \psi^*_{l-1} \cup \psi'$, the second inequality is due to $f: 2^{E}\times O^E\rightarrow \mathbb{R}_{\geq0}$ is  worst-case submodular with respect to $p(\phi)$ and $\psi'_{z, t}\subseteq \psi'_{z, |\lfloor\frac{k_z}{2}\rfloor|}$, and the equality is due to  $e_{z,t}$ maximizes the worst-case marginal utility on top of $\psi'_{z, t-1}$.  It follows that,
     \begin{eqnarray}
 \label{eq:kdd11}
 &&\sum_{e^*_l\in S^*_z} f_{wc}(e^*_l \mid \psi^*_{l-1} \cup \psi') \leq  \frac{k_z}{|\lfloor\frac{k_z}{2}\rfloor|}\sum_{t\in[|\lfloor\frac{k_z}{2}\rfloor|]}  f_{wc}(e_{z,t}\mid \psi'_{z, t-1})
  \end{eqnarray}

  We further have
       \begin{eqnarray}
 \label{eq:kdd12}
 &&\sum_{z\in[b]} \sum_{e^*_l\in S^*_z} f_{wc}(e^*_l \mid \psi^*_{l-1} \cup \psi') \leq  \sum_{z\in[b]} \frac{k_z}{|\lfloor\frac{k_z}{2}\rfloor|}\sum_{t\in[|\lfloor\frac{k_z}{2}\rfloor|]}  f_{wc}(e_{z,t}\mid \psi'_{z, t-1})
  \end{eqnarray}

  Moreover, according to  (\ref{eq:1}),  we have
         \begin{eqnarray}
 \label{eq:kdd13}\sum_{z\in[b]} \sum_{e^*_l\in S^*_z} f_{wc}(e^*_l \mid \psi^*_{l-1} \cup \psi') \geq f_{wc}(\pi^*_{wc})-f_{wc}(\pi^{mw}),
   \end{eqnarray}
  and
           \begin{eqnarray}
   &&\sum_{z\in[b]} \frac{k_z}{|\lfloor\frac{k_z}{2}\rfloor|}\sum_{t\in[|\lfloor\frac{k_z}{2}\rfloor|]}  f_{wc}(e_{z,t}\mid \psi'_{z, t-1}) \leq    \sum_{z\in[b]} (\max_{z'\in[b]}\frac{k_{z'}}{|\lfloor\frac{k_{z'}}{2}\rfloor|})\sum_{t\in[|\lfloor\frac{k_z}{2}\rfloor|]}  f_{wc}(e_{z,t}\mid \psi'_{z, t-1})\\
   &&=(\max_{z'\in[b]}\frac{k_{z'}}{|\lfloor\frac{k_{z'}}{2}\rfloor|}) \sum_{z\in[b]} \sum_{t\in[|\lfloor\frac{k_z}{2}\rfloor|]}  f_{wc}(e_{z,t}\mid \psi'_{z, t-1})\\
   &&= \frac{1}{\gamma}\sum_{z\in[b]} \sum_{t\in[|\lfloor\frac{k_z}{2}\rfloor|]}  f_{wc}(e_{z,t}\mid \psi'_{z, t-1})\leq \frac{1}{\gamma} f_{wc}(\pi^{mw})  \label{eq:kdd14}
  \end{eqnarray} where the second equality is due to (\ref{eq:patch}). (\ref{eq:kdd12}), (\ref{eq:kdd13}) and (\ref{eq:kdd14}) together imply that
  \begin{eqnarray}
  f_{wc}(\pi^*_{wc})-f_{wc}(\pi^{mw}) \leq   \frac{1}{\gamma} f_{wc}(\pi^{mw})\label{eq:kdd15}
  \end{eqnarray}

 Hence, $ f_{wc}(\pi^*_{wc}) \leq (1+\frac{1}{\gamma})f_{wc}(\pi^{mw})$.
$\Box$

\begin{lemma}
\label{lem:vijya2}
If $f: 2^{E}\times O^E\rightarrow \mathbb{R}_{\geq0}$ is adaptive monotone and adaptive submodular with respect to $p(\phi)$,  then $f_{avg}(\pi^*_{avg}) \leq  3  f_{avg}(\pi^m)$.
\end{lemma}
\emph{Proof:}  Because $f: 2^{E}\times O^E\rightarrow \mathbb{R}_{\geq0}$ is adaptive monotone, we have $f_{wc}(\pi^m)=f_{avg}(\pi^{mw}@\pi^{ma})\geq f_{avg}(\pi^{ma})$. To prove this lemma, it suffices to show that $f_{avg}(\pi^*_{avg}) \leq  3  f_{avg}(\pi^{ma})$. We next focus on proving this result.

Let $\theta$ denote a fixed run of $\pi^{ma}$. Specifically, given a fixed run $\theta$ of $\pi^{ma}$, for each $z\in[b]$ and any $t\in [\lceil\frac{k_z}{2}\rceil]$, let $\psi^\theta_{z,t}$ represent the partial realization of all items that are selected by $\pi^{ma}$ before (including) the $t$-th iteration in meta-round $z$. Moreover, let $\psi^\theta$ represent the final observation after running $\pi^{ma}$ conditioned on $\theta$. For any $\theta$, any $z\in[b]$ and any $t\in [\lceil\frac{k_z}{2}\rceil]$, let $e^\theta_{z,t}$ denote the $t$-th selected item in meta-round $z$ conditioned on $\theta$. According to the design of  $\pi^{ma}$, we have
 \begin{eqnarray}
 \label{eq:isr}
 e^\theta_{z,t} \in \arg\max_{e\in E_z} f_{avg}(e \mid \psi^\theta_{z, t-1})
 \end{eqnarray}

This, together with the facts that $f: 2^{E}\times O^E\rightarrow \mathbb{R}_{\geq0}$ is  worst-case submodular with respect to $p(\phi)$ and $ \psi^\theta_{z, t-1}\subseteq \psi^\theta$, implies the following inequality
 \begin{eqnarray}
 \label{eq:isr1}
 f_{avg}(e^\theta_{z,t} \mid \psi^\theta_{z, t-1})\geq \max_{e\in E_z} f_{avg}(e \mid \psi^\theta)
  \end{eqnarray}

Let $S^{\theta}_z \in \arg\max_{O: O\subseteq E_z, |O|\leq k_z} \sum_{e\in O} f_{avg}(e \mid \psi^\theta)$,  (\ref{eq:isr1}) implies that for any $\theta$ and $z\in[b]$,
   \begin{eqnarray}
 \label{eq:isr2}
 \sum_{t\in[\lceil\frac{k_z}{2}\rceil]} f_{avg}(e^\theta_{z,t} \mid \psi^\theta_{z, t-1})\geq \frac{\lceil\frac{k_z}{2}\rceil}{ k_z}\sum_{e\in S^{\theta}_z } f_{avg}(e \mid \psi^\theta) \geq  \frac{1}{2}\sum_{e\in S^{\theta}_z } f_{avg}(e \mid \psi^\theta)
  \end{eqnarray}

 Assume $\Lambda$ is the set of all possible runs of $\pi^{ma}$, for each possible run $\theta\in \Lambda$ of $\pi^{ma}$, let $p(\theta)=\Pr[\Theta=\theta]$ denote the probability that $\theta$ occurs, the following chain proves this lemma:
\begin{eqnarray*}
&&f_{avg}(\pi^*_{avg}) - f_{avg}(\pi^{ma})\leq\sum_{\theta\in \Lambda}p(\theta)(\sum_{z\in[b]}\sum_{e\in S^{\theta}_z } f_{avg}(e \mid \psi^\theta))\\
&&\leq  \sum_{\theta\in \Lambda}p(\theta) (\sum_{z\in[b]} \sum_{t\in[\lceil\frac{k_z}{2}\rceil]} 2\times f_{avg}(e^\theta_{z,t} \mid \psi^\theta_{z, t-1}))\\
&&=  2\times \sum_{\theta\in \Lambda}p(\theta)(\sum_{z\in[b]} \sum_{t\in[\lceil\frac{k_z}{2}\rceil]} f_{avg}(e^\theta_{z,t} \mid \psi^\theta_{z, t-1}))\\
&&= 2\times f_{avg}(\pi^{ma})
\end{eqnarray*}
The first inequality is due to $f: 2^{E}\times O^E\rightarrow \mathbb{R}_{\geq0}$ is adaptive submodular with respect to $p(\phi)$ and Lemma 5.3 from \citep{golovin2011adaptive}.  The second inequality is due to (\ref{eq:isr2}). $\Box$

\subsection{Extension to Weighted Robust Adaptive Submodular Maximization}
So far we focus on maximizing the minimum of worst-case
and average-case performance. However, in practise, it is often the case that one prioritizes average-case performance
over worst-case performance (or the opposite). To this end, in this section, we introduce a  Weighted Robust Adaptive Submodular Maximization problem and our objective is to achieve a good trade-off between worst-case and average-case performance.

Formally, define $\alpha^\beta(\pi)=\min\{\beta \frac{f_{wc}(\pi)}{f_{wc}(\pi^*_{wc})}, (1-\beta)\frac{f_{avg}(\pi)}{f_{avg}(\pi^*_{avg})}\}$ as the \emph{$\beta$-robustness ratio} of $\pi$, where the controlling parameter  $\beta\in(0,1)$ reflects the priority for   average-case performance over worst-case performance. Specifically, we can choose a larger $\beta$ if we would like to put a high priority on the average-case performance. It is easy to verify that maximizing $\alpha^\beta(\pi)$ reduces to maximizing the unweighted robustness ratio $\alpha(\pi)$ if we set $\beta=1/2$.

\subsubsection{Cardinality Constraints}
We first introduce the weighted robust adaptive submodular maximization problem subject to  cardinality constraints:
 \[\max_{\pi}\{\alpha^\beta(\pi)\mid |E(\pi, \phi)|\leq k \mbox{ for all } \phi\in U^+\}\]

Following the framework of the unweighted hybrid policy $\pi^{h}$, we design a weighted hybrid policy $\pi^{h+}$ which runs the adaptive worst-case greedy policy $\pi^g$ (Algorithm \ref{alg:LPP3}) to select $\lfloor q\times k\rfloor$ items, where $q\in[0,1]$ is a controlling parameter which will be optimized later, then runs the adaptive average-case greedy policy $\pi^{a+}$ which follows the same greedy rule as specified in Algorithm \ref{alg:LPP4} to select $\lceil(1-q)\times k\rceil$ items, ignoring the observation obtained from running $\pi^g$. Note that if we set $q=1/2$, then $\pi^{h+}$ recovers its unweighted version  $\pi^{h}$. We next optimize the selection of $q\in[0,1]$.

According to \citep{golovin2011adaptive}, if $f: 2^{E}\times O^E\rightarrow \mathbb{R}_{\geq0}$ is adaptive monotone and adaptive submodular,  then $\pi^{a+}$ achieves a $1-e^{-\frac{\lceil(1-q)\times k\rceil}{k}}$-approximation ratio for the average-case performance, i.e.,
\begin{eqnarray}
\label{eq:lastlast1}
 f_{avg}(\pi^{a+})\geq  (1-e^{-\frac{\lceil(1-q)\times k\rceil}{k}})f_{avg}(\pi^*_{avg})
 \end{eqnarray}

Hence,  $ f_{avg}(\pi^{a+})\geq  (1-e^{-(1-q)})f_{avg}(\pi^*_{avg}).$ Moreover, due to (\ref{eq:last}), $\pi^g_{\lfloor q\times k\rfloor}$ achieves a $1-e^{-\frac{\lfloor q\times k\rfloor}{k}}$-approximation ratio for the worst-case performance, i.e.,
\begin{eqnarray}
\label{eq:lastlast12}
 f_{wc}(\pi^g_{\lfloor q\times k\rfloor})\geq  (1-e^{-\frac{\lfloor q\times k\rfloor}{k}})f_{wc}(\pi^*_{wc})
 \end{eqnarray}

Hence,
 \begin{eqnarray}
 \alpha^\beta(\pi^{h+})&=&\min\{\beta \frac{f_{wc}(\pi^{h+})}{f_{wc}(\pi^*_{wc})}, (1-\beta)\frac{f_{avg}(\pi^{h+})}{f_{avg}(\pi^*_{avg})}\}\\
 &\geq& \min\{\beta \frac{f_{wc}(\pi^g_{\lfloor q\times k\rfloor})}{f_{wc}(\pi^*_{wc})}, (1-\beta)\frac{f_{avg}(\pi^{a+})}{f_{avg}(\pi^*_{avg})}\}\\
 &\geq& \min\{\beta (1-e^{-\frac{\lfloor q\times k\rfloor}{k}}), (1-\beta) (1-e^{-(1-q)})\}
 \end{eqnarray}

To achieve the best performance, we select the $q\in[0,1]$ that maximizes $ \min\{\beta (1-e^{-\frac{\lfloor q\times k\rfloor}{k}}), (1-\beta) (1-e^{-(1-q)})\}$. It turns out that the optimal $q$ is approximately $-\ln \frac{2\beta-1+\sqrt{(2\beta-1)^2+4\beta(1-\beta)/e}}{2\beta}$  when $k$ is large. It is easy to verify that for a fixed $k$, as $\beta$ increases, the optimal $q$ decreases, indicating that our policy selects more items using the  average-case greedy policy $\pi^{a+}$.

\subsubsection{Partition Matroid Constraints}
We next consider the case of matroid constraints:
 \[\max_{\pi}\{\alpha^\beta(\pi)\mid E(\pi, \phi)\in \mathcal{I} \mbox{ for all } \phi\in U^+\}\]

 We follow the framework of the unweighted hybrid policy $\pi^{m}$ to design a weighted hybrid policy $\pi^{m+}$. We first introduce two subroutines and our final policy $\pi^{m+}$ is a concatenation  of these two subroutines.

 The first subroutine $\pi^{ma+}$ is a generalized version of $\pi^{ma}$. Specifically, $\pi^{ma+}$ runs in $b$ meta-rounds, and in each meta-round $z\in[b]$, it selects   $\lceil (1-q) \times k_z\rceil$ items from $E_z$ in   $\lceil (1-q) \times k_z\rceil$ iterations such that in each iteration $t\in [\lceil (1-q) \times k_z\rceil]$ it selects an item $e_{z,t}$ that maximizes the average-case marginal utility on top of the current partial realization $\psi_{z, t-1}$: $e_{z,t} \in \argmax_{e\in E_z} f_{avg}(e\mid \psi_{z, t-1})$. Here, $q\in[0,1]$ is a controlling parameter which will be optimized later. Note that if we set $q=1/2$, then $\pi^{ma+}$ recovers its unweighted version  $\pi^{ma}$.

 The second subroutine  $\pi^{mw+}$ is a generalized version of $\pi^{mw}$. Specifically, $\pi^{mw+}$ runs in $b$ meta-rounds, and in each meta-round $z\in[b]$, it selects  $\lfloor q \times k_z\rfloor$ items from $E_z$ in   $\lfloor q \times k_z\rfloor$ iterations such that in each iteration $t\in [\lfloor q \times k_z\rfloor]$ it selects an item $e_{z,t}$ that maximizes the worst-case marginal utility on top of the current partial realization $\psi_{z, t-1}$: $e_{z,t} \in \argmax_{e\in E_z} f_{wc}(e\mid \psi_{z, t-1})$.  Note that if we set $q=1/2$, then $\pi^{mw+}$ recovers its unweighted version  $\pi^{mw}$.

 Our final policy $\pi^{m+}$ runs $\pi^{mw+}$ first, then runs  $\pi^{ma+}$, ignoring the observation obtained from running $\pi^{mw+}$. Thus, $\pi^{m+}$ can also be represented as $\pi^{mw+}@\pi^{ma+}$. We next optimize the selection of $q$.

Let $\gamma^+ = \min_{z\in[b]} \frac{\lfloor q \times k_z\rfloor}{k_z}$ and assume $\gamma^+ =(1-\epsilon)q$ for some $\epsilon \leq 1$. Note that if $\min_{z\in[b]} k_z$ is large, then $\epsilon$ approaches zero, hence, $\gamma^+$ can be approximated by $q$. Following the same proof of Lemma \ref{lem:vijya1}, except that we replace $\lceil\frac{k_z}{2}\rceil$ (resp. $\lfloor\frac{k_z}{2}\rfloor$)  by $\lceil (1-q) \times k_z\rceil$ (resp. $\lfloor q \times k_z\rfloor$) for each $z\in[b]$, we have $f_{wc}(\pi^*_{wc}) \leq  (1+\frac{1}{\gamma^+}) f_{wc}(\pi^{m+})= (1+\frac{1}{(1-\epsilon)q})f_{wc}(\pi^{m+})$. Following the same proof of Lemma \ref{lem:vijya2}, except that we replace $\lceil\frac{k_z}{2}\rceil$ (resp. $\lfloor\frac{k_z}{2}\rfloor$)  by $\lceil (1-q) \times k_z\rceil$ (resp. $\lfloor q \times k_z\rfloor$) for each $z\in[b]$,  we have $f_{avg}(\pi^*_{avg}) \leq  (1+\frac{1}{1-q}) f_{avg}(\pi^{m+})$. Hence,
\begin{eqnarray}
\alpha^\beta(\pi^{m+}) &=& \min\{\beta \frac{f_{wc}(\pi^{m+})}{f_{wc}(\pi^*_{wc})}, (1-\beta)\frac{f_{avg}(\pi^{m+})}{f_{avg}(\pi^*_{avg})}\}\\
&\geq& \min\{\beta/  (1+\frac{1}{(1-\epsilon)q}), (1-\beta)/ (1+\frac{1}{1-q})\}
 \end{eqnarray}

 To achieve the best performance, we select the $q$ that maximizes $\min\{\beta/  (1+\frac{1}{(1-\epsilon)q}), (1-\beta)/ (1+\frac{1}{1-q})\}$. It turns out that the optimal $q$ is approximately $\frac{1-\beta}{\beta+\sqrt{3\beta^2-3\beta+1}}$ when $\min_{z\in[b]} k_z$ is large. It is easy to verify that for a fixed set of integers $\{k_z\mid z\in [b]\}$, as $\beta$ increases, the optimal $q$ decreases, indicating that our policy selects more items using the  average-case greedy policy $\pi^{ma+}$.

\subsection{Applications}
\label{sec:app2}
We next introduce three applications whose utility function satisfies all four conditions stated in Theorem \ref{thm:4}, i.e., worst-case monotonicity, worst-case submodularity, adaptive submodularity and minimal dependency.
\paragraph{Pool-based Active Learning.} In Section \ref{sec:app}, we have shown that the utility function $f: 2^{E}\times O^E\rightarrow \mathbb{R}_{\geq0}$ of pool-based active learning  is worst-case monotone, worst-case submodular with respect to $p(\phi)$ and it satisfies the property of minimal dependency. \cite{golovin2011adaptive} have shown that  $f: 2^{E}\times O^E\rightarrow \mathbb{R}_{\geq0}$ of pool-based active learning  is adaptive  submodular with respect to $p(\phi)$. Thus, we have the following proposition.

\begin{proposition}
The utility function $f: 2^{E}\times O^E\rightarrow \mathbb{R}_{\geq0}$ of pool-based active learning is worst-case monotone, worst-case submodular, adaptive submodular with respect to $p(\phi)$ and it satisfies the property of minimal dependency.
\end{proposition}

 \paragraph{The Case when Items are Independent.} In Section \ref{sec:app}, we have shown that  the utility function $f: 2^{E}\times O^E\rightarrow \mathbb{R}_{\geq0}$  is worst-case submodular with respect to $p(\phi)$ if three conditions are satisfied. Because the first condition, i.e., for all $e\in E$ and all possible partial realizations $\psi$ such that $e\notin \mathrm{dom}(\psi)$, $ O(e, \psi)=O(e, \emptyset)$, is satisfied if the states of all items are independent of each other. We conclude that $f: 2^{E}\times O^E\rightarrow \mathbb{R}_{\geq0}$  is worst-case submodular with respect to $p(\phi)$ if the following three conditions are satisfied:
 \begin{enumerate}
  \item  The states of items are independent of each other, i.e., for all $e\in E$ and $\psi$ such that $e\notin \mathrm{dom}(\psi)$, $\Pr[\phi(e)=o \mid \phi\sim \psi]=  \Pr[\phi(e)=o]$.
   \item $f: 2^{E}\times O^E\rightarrow \mathbb{R}_{\geq0}$ is pointwise submodular and pointwise monotone with respect to $p(\phi)$.
   \item $f: 2^{E}\times O^E\rightarrow \mathbb{R}_{\geq0}$ satisfies the property of minimal dependency.
\end{enumerate}
 Moreover, \cite{golovin2011adaptive} have shown that  $f: 2^{E}\times O^E\rightarrow \mathbb{R}_{\geq0}$ is adaptive submodular if the above three conditions are satisfied. Thus, we have the following proposition.
\begin{proposition}
If the above three conditions are satisfied, then $f: 2^{E}\times O^E\rightarrow \mathbb{R}_{\geq0}$ is worst-case monotone, worst-case submodular, adaptive submodular with respect to $p(\phi)$ and it satisfies the property of minimal dependency.
\end{proposition}

 \paragraph{Adaptive Viral Marketing.} In Section \ref{sec:app}, we have shown that the utility function $f: 2^{E}\times O^E\rightarrow \mathbb{R}_{\geq0}$ of adaptive viral marketing is worst-case monotone, worst-case submodular with respect to $p(\phi)$ and it satisfies the property of minimal dependency. \cite{golovin2011adaptive} have shown that  $f: 2^{E}\times O^E\rightarrow \mathbb{R}_{\geq0}$ of adaptive viral marketing  is adaptive submodular. Thus, we have the following proposition.
 \begin{proposition}
\label{pro:13}
The utility function $f: 2^{E}\times O^E\rightarrow \mathbb{R}_{\geq0}$ of adaptive viral marketing is worst-case monotone, worst-case submodular, adaptive submodular with respect to $p(\phi)$ and it satisfies the property of minimal dependency.
\end{proposition}

\section{Performance Evaluation}

We conduct experiments to evaluate the average-case performance of our proposed adaptive algorithms: \emph{Average-case  Policy $\pi^a$} (AP), \emph{Worst-case Policy $\pi^g$} (WP) and \emph{Hybrid Policy $\pi^h$} (HP) subject to cardinality constraints.  Our goal is to investigate the price of ``robustness'' by comparing the average-case performance of AP with that of WP/HP. We perform our evaluations in the context of pool-based active learning. Please refer to Section \ref{sec:activelearning} for a detailed description of this application. Recall that (\ref{eq:10}) measures the reduction in version space mass. Our goal is to sequentially query the labels of (a.k.a. select) at most $k$ data points, each selection is based on the observation of previously obtained labels, to maximize the reduction in version
space mass.
We consider a group of hypotheses $\mathcal{H}$ with $50$ unlabeled data points.  For each hypothesis $h\in \mathcal{H}$, its probability is decided as $\frac{q_h}{\sum_{h'\in \mathcal{H}} q_{h'}}$, where each $q_{h}$ is sampled from $(0,1)$ uniformly at random.  The realized label of each data point is selected from a  set of possible labels.  In our experiments, we vary the cardinality constraint $k$, the number of hypotheses $|\mathcal{H}|$, and the size of the label set, and present the average-case performance of all algorithms. The datasets can be downloaded from \url{https://www.dropbox.com/s/4v2i9darmy7t1dv/data-robust.zip?dl=0}.
\begin{figure*}[hptb]
\hspace*{-0.75in}
\includegraphics[scale=0.2]{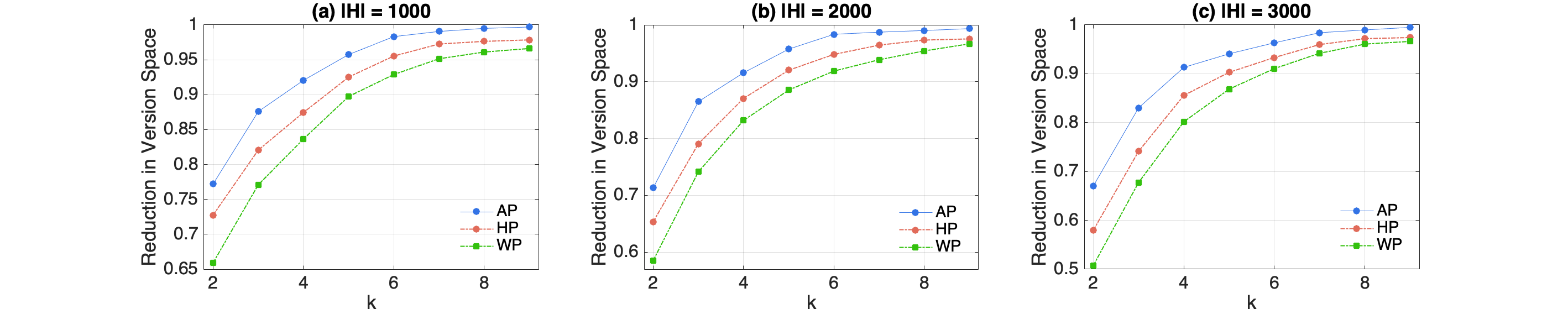}
\caption{Reduction in version space mass vs. cardinality constraint $k$.}
\label{fig:reduction_k}
\end{figure*}

\begin{figure*}[hptb]
\hspace*{-0.75in}
\includegraphics[scale=0.2]{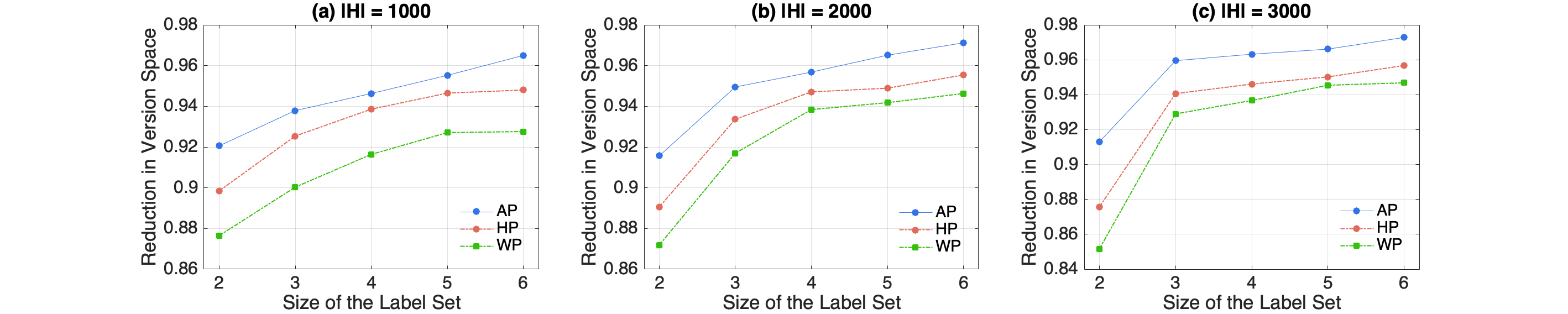}
\caption{Reduction in version space mass vs. number of possible labels for each data point.}
\label{fig:reduction_label_size}
\end{figure*}
Our first set of experiments evaluate the average-case performance of the algorithms as measured by the yielded reduction in version space mass with respect to the changes in the value of $k$. The results are plotted in Figure \ref{fig:reduction_k}. In this experiment, we consider binary data points, i.e., each data point has two possible labels. As shown in the figure, the $x$-axis refers to the value of $k$, ranging from two to nine. The $y$-axis refers to the reduction in version space mass produced by the corresponding algorithms. Figure \ref{fig:reduction_k}(a) shows the results where $1,000$ hypotheses are considered. We observe that as the value of $k$ increases, the reduction in version space mass increases for all algorithms. Intuitively a larger $k$ indicates that more data points can be selected in the output, leading to a higher reduction in version space mass. We also observe that AP achieves the best performance while HP outperforms WP. In particular, AP outperforms WP by up to $17.11\%$, and AP outperforms HP by up to $6.10\%$. Figure \ref{fig:reduction_k}(b) shows the results where $2,000$ hypotheses are considered.  Similarly, we observe that AP outperforms HP and WP, and HP outperforms WP. In particular, AP outperforms WP by up to $21.85\%$, and AP outperforms HP by up to $9.14\%$.  Figure \ref{fig:reduction_k}(c) shows the results where $3,000$ hypotheses are considered.  Again, we observe that AP outperforms HP and WP, and HP outperforms WP. In particular, AP outperforms WP by up to $32.03\%$, and AP outperforms HP by up to $15.63\%$. These results verify the superiority of HP in striking a balance between worst-case performance and average-case performance. We also notice that these gaps become smaller as $k$ increases. This observation can be partially explained by the property of diminishing returns of our utility function, i.e., the gain by selecting a new data point on top of a larger set is no greater than the marginal utility gained by selecting the same data point on top of a smaller set.

Our second set of experiments explore the impact of the size of label set on the reduction in version space mass, as illustrated in Figure \ref{fig:reduction_label_size}. We set the cardinality constraint $k=4$. The $x$-axis refers to the size of the label set, ranging from two to six. The $y$-axis refers to the reduction in version space mass produced by the corresponding algorithms.  Figure \ref{fig:reduction_label_size}(a) shows the results where $1,000$ hypotheses are considered. We observe that as the size of the label set increases, the reduction in version space mass increases for all algorithms. We also observe that AP outperforms WP by up to $5.02\%$, and AP outperforms HP by up to $2.46\%$. Figure \ref{fig:reduction_label_size}(b) shows the results where $2,000$ hypotheses are considered. AP outperforms WP by up to $5.03\%$, and AP outperforms HP by up to $2.82\%$. Figure \ref{fig:reduction_label_size}(c) shows the results where $3,000$ hypotheses are considered. AP outperforms WP by up to $7.19\%$, and AP outperforms HP by up to $4.26\%$.

\begin{figure*}[hptb]
\hspace*{-0.75in}
\includegraphics[scale=0.2]{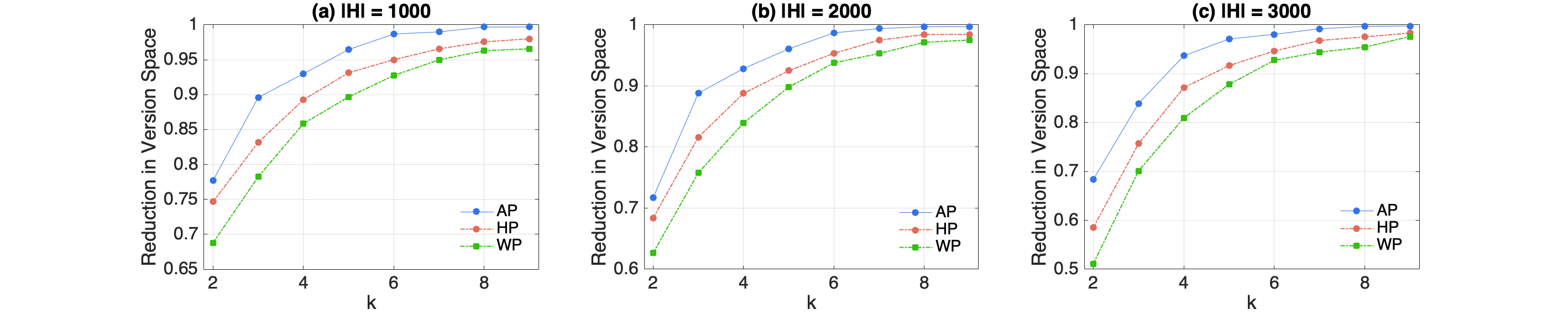}
\caption{Reduction in version space mass vs. cardinality constraint $k$.}
\label{fig:reduction_k_hybrid}
\end{figure*}

In Figure \ref{fig:reduction_k_hybrid}, we consider the scenario where data points have various numbers of possible labels. We randomly divide our $50$ unlabeled data points into three groups. The first group contains $40$ data points with binary labels. The second group contains $5$ data points with three possible labels. The third group contains $5$ data points with four possible labels. As shown in the figure, the $x$-axis holds the value of $k$, ranging from two to nine. The $y$-axis holds the reduction in version space mass generated by the corresponding algorithms. We observe that as expected, the reduction in version space mass increases as the value of $k$ increases. Figure \ref{fig:reduction_k_hybrid}(a) shows the results where $1,000$ hypotheses are considered.  AP outperforms WP by up to $12.97\%$, and AP outperforms HP by up to $4.07\%$. Figure \ref{fig:reduction_k_hybrid}(b) shows the results where $2,000$ hypotheses are considered. AP outperforms WP by up to $14.41\%$, and AP outperforms HP by up to $4.85\%$.  Figure \ref{fig:reduction_k_hybrid}(c) shows the results where $3,000$ hypotheses are considered.  AP outperforms WP by up to $33.71\%$, and AP outperforms HP by up to $16.86\%$.  Moreover, these gaps become smaller as $k$ increases.

\section{Conclusion}
In this paper, we study two variants of adaptive submodular maximization problems. We first develop a policy for maximizing a worst-case submodular function subject to a $p$-system constraint in worst case setting. Then we develop hybrid policies that achieve good performances in both average case setting and worst case setting subject to cardinality constraints and partition matroid constraints. In the future, we would like to study the robust adaptive submodular maximization problem subject to more general constraints such as $p$-system constraints.
\bibliographystyle{ijocv081}
\bibliography{reference}

\clearpage

\setcounter{page}{1}
\begin{APPENDICES}
\section{Missing Definitions, Lemmas and Proofs}
\subsection{Proof of Theorem \ref{thm:important}}
\emph{Proof:}  The proof of the running time is trivial. Because $\pi^f$ takes $k$ rounds to select $k$ items, and the running time of each round is bounded by $\frac{n}{k}\log\frac{1}{\epsilon}$ which is the size of the random set $H$. Thus, the total running time is bounded by $O(k\times \frac{n}{k}\log\frac{1}{\epsilon})=O(n\log\frac{1}{\epsilon})$.

We next prove the approximation ratio of $\pi^f$. Let $\phi'$ denote the worst-case realization for $\pi^f$, i.e., $\phi'=\argmin_{\phi}\mathbb{E}[f(E(\pi^f, \phi), \phi)]$. Let random variables $\Psi'_{t}$ denote the partial realization after selecting the first $t$ items $\mathbf{S}_{t}$, which is also a random variable, conditioned on $\phi'$. We can represent the expected worst-case utility $\tilde{f}_{wc}(\pi^f)$ of $\pi^f$ as follows:
\begin{eqnarray}
&&\tilde{f}_{wc}(\pi^f)=\mathbb{E}_{\Pi}[f(E(\pi^f, \phi'), \phi')]~\nonumber\\
&&= \sum_{t\in [k]}\{\mathbb{E}_{\Pi}[f(E(\pi^f_t, \phi'), \phi')]-\mathbb{E}_{\Pi}[f(E(\pi^f_{t-1}, \phi'), \phi')]\}~\nonumber\\
&&= \sum_{t\in [k]} \{\mathbb{E}_{\mathbf{S}_{t}, \Psi'_{t}}[f(\mathbf{S}_{t}, \Psi'_{t})]-\mathbb{E}_{\mathbf{S}_{t-1}, \Psi'_{t-1}}[f(\mathbf{S}_{t-1}, \Psi'_{t-1})]\} \label{eq:ha}
\end{eqnarray}
To prove this theorem, it suffices to show that for all $t\in [k]$,
\begin{eqnarray}
\label{eq:g}
 \mathbb{E}_{\mathbf{S}_{t}, \Psi'_{t}}[f(\mathbf{S}_{t}, \Psi'_{t})]-\mathbb{E}_{\mathbf{S}_{t-1}, \Psi'_{t-1}}[f(\mathbf{S}_{t-1}, \Psi'_{t-1})] \geq (1-\epsilon)\frac{ f_{wc}(\pi^*_{wc})-\mathbb{E}_{\mathbf{S}_{t-1}, \Psi'_{t-1}}[f(\mathbf{S}_{t-1}, \Psi'_{t-1})]}{k}
\end{eqnarray}
This is because by induction,  (\ref{eq:g}) and (\ref{eq:ha}) imply  that $\tilde{f}_{wc}(\pi^f) \geq (1-(1-\frac{1-\epsilon}{k})^k)f_{wc}(\pi^*_{wc})\geq (1-e^{-(1-\epsilon)})f_{wc}(\pi^*_{wc}) \geq (1-1/e-\epsilon)f_{wc}(\pi^*_{wc})$.

The rest of the proof is devoted to proving (\ref{eq:g}). Consider a given partial realization $\psi'_{t-1}$ after selecting the first $t-1$ items $S_{t-1}$. Let $W\in \argmax_{S\subseteq E: |S|=k} \sum_{e\in S}f_{wc}(e\mid \psi'_{t-1})$ denote the top $k$ items that have the largest worst-case marginal utility on top of $\psi'_{t-1}$. Recall that $H$ is a random set of size $\frac{n}{k}\log\frac{1}{\epsilon}$ sampled uniformly at random  from $E$, we first show that the probability that $H \cap W =\emptyset$ is upper bounded by $e^{-|H|\frac{k}{n}}$.
\begin{eqnarray*}
\Pr[H \cap W = \emptyset]&\leq&{(1-\frac{|W|}{|E|})}^{|H|} ={(1-\frac{k}{n})}^{|H|}\leq e^{-|H|\frac{k}{n}}
\end{eqnarray*}

It follows that $\Pr[H \cap W \neq \emptyset]\geq 1- e^{-\frac{|H|}{n}k}$. Because we assume $|H|=\frac{n}{k}\log\frac{1}{\epsilon}$, it follows that \begin{equation}\label{eq:hoho}\Pr[H \cap W\neq \emptyset]\geq 1-e^{-|H|\frac{k}{n}} \geq 1-  e^{-\frac{\frac{n}{k}\log\frac{1}{\epsilon}}{n}k}\geq 1-\epsilon
\end{equation}

Now we are ready to prove (\ref{eq:g}). Again, consider a fixed partial realization $\psi'_{t-1}$ after selecting the first $t-1$ items $S_{t-1}$, we have
\begin{eqnarray}
&&\mathbb{E}_{\mathbf{S}_{t}, \Psi'_{t}}[f(\mathbf{S}_{t}, \Psi'_{t})\mid S_{t-1}, \psi'_{t-1}]-f(S_{t-1}, \psi'_{t-1})~\nonumber\\
&&=  \mathbb{E}_{\mathbf{e}'_t}[f(S_{t-1}\cup \{\mathbf{e}'_t\}, \psi'_{t-1}\cup\{(\mathbf{e}'_t, \phi'(\mathbf{e}'_t))\})-f(S_{t-1}, \psi'_{t-1})\mid S_{t-1}, \psi'_{t-1}]~\nonumber\\
&&\geq  \mathbb{E}_{\mathbf{e}'_t}[ f_{wc}(\mathbf{e}'_t \mid \psi'_{t-1}) \mid S_{t-1}, \psi'_{t-1}]\geq \mathbb{E}_{H}[\max_{e\in H } f_{wc}(e\mid \psi'_{t-1})\mid S_{t-1}, \psi'_{t-1}]~\nonumber\\
&&\geq \Pr[H \cap W \neq \emptyset] \frac{\sum_{e\in H}f_{wc}(e\mid \psi'_{t-1})}{k}\geq (1-\epsilon)\frac{\sum_{e\in W}f_{wc}(e\mid \psi'_{t-1})}{k}\geq (1-\epsilon)\frac{\sum_{i\in [k]}f_{wc}(e^*_i\mid \psi'_{t-1})}{k}~\nonumber\\
&&\geq (1-\epsilon)\frac{\sum_{i\in [k]}f_{wc}(e^*_i\mid \psi'_{t-1}\cup\psi^*_{i-1})}{k}=(1-\epsilon)\frac{ f(S_{t-1}\cup S^*_k, \psi'_{t-1}\cup\psi^*_{k-1})-f(S_{t-1}, \psi'_{t-1})}{k}~\nonumber\\
&&\geq (1-\epsilon) \frac{ f( S^*_k, \psi^*_{k})-f(S_{t-1}, \psi'_{t-1})}{k}~\nonumber\\
&&=  (1-\epsilon) \frac{f_{wc}(\pi^*_{wc})-f(S_{t-1}, \psi'_{t-1})}{k}\label{eq:h}
\end{eqnarray}
The third inequality is due to each item of $W$ has equal probability of being included in $H \cap W$. The fourth inequality is due to (\ref{eq:hoho}). The fifth  inequality is due to $W\in \argmax_{S\subseteq E: |S|=k} \sum_{e\in S}f_{wc}(e\mid \psi'_{t-1})$. The sixth inequality is due to (\ref{eq:1}).
Taking the expectation of (\ref{eq:h}) over $(\mathbf{S}_{t-1}, \Psi'_{t})$, we have (\ref{eq:g}). $\Box$

\subsection{Proof of Proposition \ref{pro:1}}
\emph{Proof:} Consider any two partial realizations $\psi$ and $\psi'$ such that $\psi\subseteq \psi'$ and any item $e\in E\setminus \mathrm{dom}(\psi')$, we have
\begin{eqnarray}
f_{wc}(e\mid \psi)&& = \min_{o \in O(e, \psi)}f(\mathrm{dom}(\psi)\cup\{e\}, \psi\cup\{(e, o)\})-f(\mathrm{dom}(\psi), \psi)~\nonumber\\
&& = \min_{o \in O(e, \psi)} 1- p_H(\mathcal{H}(\psi\cup\{(e, o)\}))-(1- p_H(\mathcal{H}(\psi)))~\nonumber\\
&&= \min_{o \in O(e, \psi)}  p_H(\mathcal{H}(\psi)) - p_H(\mathcal{H}(\psi\cup\{(e, o)\}))~\nonumber
\end{eqnarray}
Similarly, we have
\begin{eqnarray}
f_{wc}(e\mid \psi')&& =  \min_{o \in O(e, \psi')}  p_H(\mathcal{H}(\psi')) - p_H(\mathcal{H}(\psi'\cup\{(e, o)\}))~\nonumber
\end{eqnarray}
We next show that $f_{wc}(e\mid \psi)\geq f_{wc}(e\mid \psi')$. We first consider the case when there exists a state $o' \in \argmin_{o \in O(e, \psi)} p_H(\mathcal{H}(\psi)) - p_H(\mathcal{H}(\psi\cup\{(e, o)\}))$ such that $o'\in O(e, \psi')$. In this case, we have
\begin{eqnarray}
\label{eq:a}
f_{wc}(e\mid \psi)= p_H(\mathcal{H}(\psi)) - p_H(\mathcal{H}(\psi\cup\{(e, o')\}))
\end{eqnarray}
and
\begin{eqnarray}
f_{wc}(e\mid \psi')&& =  \min_{o \in O(e, \psi')}  p_H(\mathcal{H}(\psi')) - p_H(\mathcal{H}(\psi'\cup\{(e, o)\}))~\nonumber\\
&& \leq   p_H(\mathcal{H}(\psi')) - p_H(\mathcal{H}(\psi'\cup\{(e, o')\}))\label{eq:b}
\end{eqnarray}
Because $\psi\subseteq \psi'$, any hypotheses $h\in \mathcal{H}(\psi')$ that is not consistent with $\psi'\cup\{(e, o')\}$ must satisfy that  $h\in \mathcal{H}(\psi)$ and $h$ is not consistent with $\psi\cup\{(e, o')\}$. Thus, we have $\mathcal{H}(\psi')\setminus \mathcal{H}(\psi'\cup\{(e, o')\})\subseteq \mathcal{H}(\psi)\setminus \mathcal{H}(\psi\cup\{(e, o')\})$. Thus, $ p_H(\mathcal{H}(\psi)) - p_H(\mathcal{H}(\psi\cup\{(e, o')\}))\leq p_H(\mathcal{H}(\psi)) - p_H(\mathcal{H}(\psi\cup\{(e, o')\}))$. This together with (\ref{eq:a}) and (\ref{eq:b}) implies that $f_{wc}(e\mid \psi)\geq f_{wc}(e\mid \psi')$.

We next consider the case when there does not exist a state $o' \in \argmin_{o \in O(e, \psi)} p_H(\mathcal{H}(\psi)) - p_H(\mathcal{H}(\psi\cup\{(e, o)\}))$ such that $o'\in O(e, \psi')$. In this case, for any state $o'' \in \argmin_{o \in O(e, \psi)} p_H(\mathcal{H}(\psi)) - p_H(\mathcal{H}(\psi\cup\{(e, o)\}))$, we have \begin{eqnarray}
\label{eq:c}
\mathcal{H}(\psi')\subseteq \mathcal{H}(\psi)\setminus \mathcal{H}(\psi\cup\{(e, o'')\})
 \end{eqnarray}
 This is because none of the hypotheses from $\mathcal{H}(\psi')$ is consistent with $\{(e, o'')\}$. It follows that
 \begin{eqnarray}
 f_{wc}(e\mid \psi)&&= p_H(\mathcal{H}(\psi)) - p_H(\mathcal{H}(\psi\cup\{(e, o'')\}))~\nonumber\\
 &&\geq  p_H(\mathcal{H}(\psi'))~\nonumber \\
 &&\geq \min_{o \in O(e, \psi')}  p_H(\mathcal{H}(\psi')) - p_H(\mathcal{H}(\psi'\cup\{(e, o)\}))~\nonumber\\
 &&=f_{wc}(e\mid \psi')~\nonumber
 \end{eqnarray}
 The first equality is due to the definition of $o''$ and the first inequality is due to (\ref{eq:c}). This finishes the proof of this proposition. $\Box$
\subsection{Proof of Proposition \ref{pro:2}}
\emph{Proof:} The proof of worst-case monotonicity is trivial. For any $e\in E$ and $\psi$ such that $e\notin \mathrm{dom}(\psi)$, let $o'\in \argmin_{o \in O(e, \psi)}f(\mathrm{dom}(\psi)\cup\{e\}, \psi\cup\{(e, o)\})-f(\mathrm{dom}(\psi)$. Consider any realization $\phi'$ that is consistent with $\psi\cup\{(e, o')\}$, i.e., $\phi'\sim \psi\cup\{(e, o')\}$, we have
\begin{eqnarray}
f_{wc}(e \mid \psi)&&=f(\mathrm{dom}(\psi)\cup\{e\}, \psi\cup\{(e, o')\})-f(\mathrm{dom}(\psi), \psi)~\nonumber\\
&&= f(\mathrm{dom}(\psi)\cup\{e\}, \phi'\})-f(\mathrm{dom}(\psi), \phi')~\nonumber\\
&&\geq0~\nonumber
\end{eqnarray}
 The second equality is due to $f: 2^{E}\times O^E\rightarrow \mathbb{R}_{\geq0}$ satisfies the property of minimal dependency and the inequality is due to $f: 2^{E}\times O^E\rightarrow \mathbb{R}_{\geq0}$ is pointwise monotone with respect to $p(\phi)$.

We next prove that $f: 2^{E}\times O^E\rightarrow \mathbb{R}_{\geq0}$ is  worst-case submodular  with respect to $p(\phi)$. Consider any two partial realizations $\psi$ and $\psi'$ such that $\psi\subseteq \psi'$ and any item $e\in E\setminus \mathrm{dom}(\psi')$. Because of the first condition, we have $O(e, \psi) =  O(e, \psi')$. 
Then for any $o\in O(e, \psi)$, $\phi\sim \psi\cup\{(e, o)\}$ and $\phi'\sim \psi'\cup\{(e, o)\}$, we have \begin{eqnarray}
&&f(\mathrm{dom}(\psi)\cup\{e\}, \psi\cup\{(e, o)\})-f(\mathrm{dom}(\psi), \psi)~\nonumber\\
&& = f(\mathrm{dom}(\psi)\cup\{e\}, \phi)-f(\mathrm{dom}(\psi), \phi)~\nonumber\\
 &&= f(\mathrm{dom}(\psi)\cup\{e\}, \phi')-f(\mathrm{dom}(\psi), \phi')~\nonumber\\
&& \geq f(\mathrm{dom}(\psi')\cup\{e\}, \phi')-f(\mathrm{dom}(\psi'), \phi')~\nonumber\\
&&= f(\mathrm{dom}(\psi')\cup\{e\}, \psi'\cup\{(e, o)\})-f(\mathrm{dom}(\psi'), \psi')\label{eq:d}\end{eqnarray}
All equalities are due to $f: 2^{E}\times O^E\rightarrow \mathbb{R}_{\geq0}$ satisfies the property of minimal dependency and the inequality is due to  $f: 2^{E}\times O^E\rightarrow \mathbb{R}_{\geq0}$ is pointwise submodular.

Let $o'\in \argmin_{o\in  O(e, \psi)}f(\mathrm{dom}(\psi)\cup\{e\}, \psi\cup\{(e, o)\})-f(\mathrm{dom}(\psi), \psi)$, we have
\begin{eqnarray}
 f_{wc}(e\mid \psi)&& = f(\mathrm{dom}(\psi)\cup\{e\}, \psi\cup\{(e, o')\})-f(\mathrm{dom}(\psi), \psi)~\nonumber  \\
&&\geq f(\mathrm{dom}(\psi')\cup\{e\}, \psi\cup\{(e, o')\})-f(\mathrm{dom}(\psi'), \psi')~\nonumber\\
&& \geq \min_{o\in  O(e, \psi')}f(\mathrm{dom}(\psi')\cup\{e\}, \psi\cup\{(e, o)\})-f(\mathrm{dom}(\psi'), \psi')~\nonumber\\
&& = f_{wc}(e\mid \psi')
\end{eqnarray}
The first inequality is due to (\ref{eq:d}). This finishes the proof of this proposition. $\Box$
\subsection{Proof of Proposition \ref{pro:3}}
\emph{Proof:} It is easy to verify that $f: 2^{E}\times O^E\rightarrow \mathbb{R}_{\geq0}$ is worst-case monotone and it satisfies the property of minimal dependency. We next show that it is also worst-case submodular with respect to $p(\phi)$.

 Consider any two partial realizations $\psi$ and $\psi'$ such that $\psi\subseteq \psi'$ and any node $e\in E\setminus \mathrm{dom}(\psi')$. Let $Z(\psi)$ denote the set of edges whose status is observed under $\psi$ and let $E(\psi)$ denote the set of nodes that are influenced under $\psi$. Clearly, $Z(\psi) \subseteq Z(\psi')$ and $E(\psi)\subseteq E(\psi')$. Now consider the worst-case marginal utility  $f_{wc}(e\mid \psi)$ of $e$ on top of $\psi$, the worst-case realization of $e$, say $o$, occurs when  all edges from $\{(u,v)\in Z\setminus Z(\psi) \mid p(u,v)<1\}$ are blocked. Similarly, for  the worst-case marginal utility  $f_{wc}(e\mid \psi')$ of $e$ on top of $\psi'$, the worst-case realization of $e$, say $o'$, occurs when  all edges from $\{(u,v)\in Z\setminus Z(\psi') \mid p(u,v)<1\}$ are blocked. Because $Z(\psi) \subseteq Z(\psi')$ , we have $Z\setminus Z(\psi') \subseteq Z\setminus Z(\psi)$. This together with $E(\psi)\subseteq E(\psi')$ implies that  if a node $w\in E\setminus E(\psi')$ can be reached by some live edge under $(e, o')$, it must be the case that $w\in E\setminus E(\psi)$ and $w$  can be reached by some live edge under $(e, o)$. We conclude that the additional nodes influenced by $e$ on top of $\psi$ under the worst-case realization is a superset of the additional nodes influenced by $e$ on top of $\psi'$ under the worst-case realization. It follows that $f_{wc}(e\mid \psi) \geq f_{wc}(e\mid \psi')$. This finishes the proof of this proposition. $\Box$
\end{APPENDICES}

\end{document}